\RequirePackage{iftex}
\ifPDFTeX\else
  \errmessage{Use pdfLaTeX for this IEEEtran submission; XeLaTeX/LuaLaTeX will change the fonts}
\fi
\documentclass[conference]{IEEEtran}
\IEEEoverridecommandlockouts

\usepackage[utf8]{inputenc}
\usepackage{cite}
\usepackage{amsmath,amssymb,amsfonts,amsthm}
\usepackage{algorithm}
\usepackage{algpseudocode}
\usepackage{graphicx}
\usepackage{textcomp}
\usepackage{xcolor,colortbl}
\usepackage{booktabs}
\usepackage{multirow}
\usepackage{bm}
\usepackage{enumitem}
\usepackage{threeparttable}
\usepackage{subfig}
\usepackage{tikz}
\usepackage{url}
\usepackage{balance}
\usepackage[hidelinks]{hyperref}
\usepackage{cleveref}

\usepackage{amsmath,amsfonts,bm}
\def\eqref#1{equation~\ref{#1}}
\def\1{\bm{1}}

\DeclareMathAlphabet{\mathsfit}{\encodingdefault}{\sfdefault}{m}{sl}
\SetMathAlphabet{\mathsfit}{bold}{\encodingdefault}{\sfdefault}{bx}{n}

\def\gF{{\mathcal{F}}}
\def\gG{{\mathcal{G}}}

\def\chi{{\mathcal{X}}}

\newcommand{\E}{\mathbb{E}}

\newcommand{\R}{\mathbb{R}}

\graphicspath{{figs/}{./figs/}}
\newtheorem{theorem}{Theorem}[section]
\newtheorem{lemma}[theorem]{Lemma}

\newcommand{\B}{\{0, 1\}}
\newcommand{\I}{\mathbb{I}}
\newcommand{\param}{\theta}
\newcommand{\logits}{\hat{\theta}}
\newcommand{\blogits}{\boldsymbol{\hat{\theta}}}
\newcommand{\prob}{p(\cdot | \param)}
\newcommand{\data}{x}
\newcommand{\noise}{\epsilon}
\newcommand{\xh}{\tilde{\data}_{hard}}
\newcommand{\xs}{\tilde{\data}_{soft}}

\definecolor{mypurple}{HTML}{b05f8c}
\definecolor{mypink}{HTML}{f58297}
\definecolor{myyellow}{HTML}{f7d35c}
\definecolor{lightpink}{HTML}{FFBFBF}
\definecolor{lightyellow}{HTML}{FFEF93}
\definecolor{lightpurple}{HTML}{E4C1EA}
\definecolor{lightgreen}{HTML}{73B405}
\definecolor{myorange}{RGB}{244,106,18}
\definecolor{myblue}{RGB}{0,111,190}
\definecolor{mygreen}{RGB}{0,127,128}
\definecolor{myred}{RGB}{228,46,36}
\definecolor{mydark}{RGB}{114,44,114}
\definecolor{mymiddle}{RGB}{144,44,144}
\definecolor{mylight}{RGB}{167,44,167}
\definecolor{nvidia}{HTML}{73B405}
\newcommand{\new}[1]{\textcolor{black}{#1}}
\newcommand{\scriptbadge}[1]{%
  \tikz[baseline=(char.base)]\node[
    circle,
    fill=black,
    text=white,
    font=\sffamily\bfseries\fontsize{6.5}{6.5}\selectfont,
    inner sep=0.2pt,
    minimum size=1.28em
  ] (char) {#1};%
}
\newcommand{\scriptlink}[2]{\hyperlink{#1}{\scriptbadge{#2}}}

\Crefformat{figure}{Fig.~#2#1#3}
\Crefname{subfigure}{Fig.}{Figs.}
\Crefname{figure}{Fig.}{Figs.}
\Crefformat{table}{Table~#2#1#3}
\Crefname{appendix}{Appendix}{Appendices}
\Crefname{section}{Section}{Sections}

\def\BibTeX{{\rm B\kern-.05em{\sc i\kern-.025em b}\kern-.08em
    T\kern-.1667em\lower.7ex\hbox{E}\kern-.125emX}}

\begin{document}

\title{DEFT: Differentiable Automatic Test Pattern Generation}

\author{\IEEEauthorblockN{Wei Li, Yang Zou, Yixin Liang, Jos\'{e} Moura, and Shawn Blanton}
\IEEEauthorblockA{\textit{Carnegie Mellon University}\\
Pittsburgh, PA, USA}}

\maketitle

\begin{abstract}
\label{sec:abstract}
Modern IC complexity drives test pattern growth, 
with the majority of patterns targeting a small set of hard-to-detect (HTD) faults.
This motivates new ATPG algorithms to improve test effectiveness specifically for HTD faults.
This paper presents DEFT (Differentiable Automatic Test Pattern Generation), 
a new ATPG approach that reformulates the discrete ATPG problem as a continuous optimization task. 
DEFT introduces a mathematically grounded reparameterization that aligns the expected continuous objective with discrete fault-detection semantics, enabling reliable gradient-based pattern generation.
To ensure scalability and stability on deep circuit graphs, DEFT integrates a custom CUDA kernel for efficient forward-backward propagation and applies gradient normalization to mitigate vanishing gradients.
Compared to a leading commercial tool on a wide range of benchmarks, 
DEFT \textbf{reduced the pattern count by 27.3\%} on average and by up to 75.9\%.
DEFT also supports practical ATPG settings such as partial assignment pattern generation, producing patterns 
with \textbf{19.3\% fewer 0/1 bits} while still \textbf{detecting 35\% more faults}.
These results indicate DEFT is a promising and effective ATPG engine, offering a valuable complement to existing heuristics.
\end{abstract}


\section{Introduction}
\label{sec:intro}

Pattern sets are growing due to the increasing circuit size \cite{HIR2024_TestTech}
and the adoption of fault models with much larger fault universes
(cell-aware \cite{hapke2014cell}, PEPR \cite{li2022pepr}, etc.). 
A large portion of these patterns is dedicated to only a small set of hard-to-detect (HTD) faults.
For instance, on the NVDLA benchmark, 71.8\% of patterns are generated for the last 1\% of faults in a MAC cell.
\new{Commercial tools are already highly efficient at detecting HTD faults in terms of runtime, 
yet our experiments indicate that the resulting pattern count remains far from optimal even under the 
maximum compaction-effort setting, which disproportionately inflates test-set size and test cost.}
This motivates new ATPG approaches that more effectively target HTD faults, curb pattern growth, and scale to large designs.

Two main avenues have been explored, each with critical limitations: 1) AI-driven Heuristics: 
Methods using reinforcement learning \cite{li2024smartatpg} seek to improve the decision-making of traditional heuristic solvers. 
These approaches face significant scalability hurdles. 
% Replacing a highly-optimized, few-cycle decision kernel with a complex neural network inference at every step of the search is often computationally impractical 
% on large designs. 
2) SAT-based Methods: 
Reformulating ATPG as a Boolean Satisfiability (SAT) problem is highly effective for proving fault redundancy.
However, this approach suffers from a mismatch: 
SAT solvers are optimized for decision problems (``is this fault testable?"), 
whereas ATPG has evolved into an optimization problem (``find a minimal pattern set to maximize the fault coverage"). 
While extendable, SAT's efficiency drops precipitously when applied to large-scale test compression and compaction, 
often resulting in an increase in the number of tests required to achieve desired levels of fault coverage \cite{chao2025pastatpg}.
% and an inability to perform the global, multi-target compaction that heuristic tools excel at.

\begin{figure*}
    \centering
    \includegraphics[width=\textwidth]{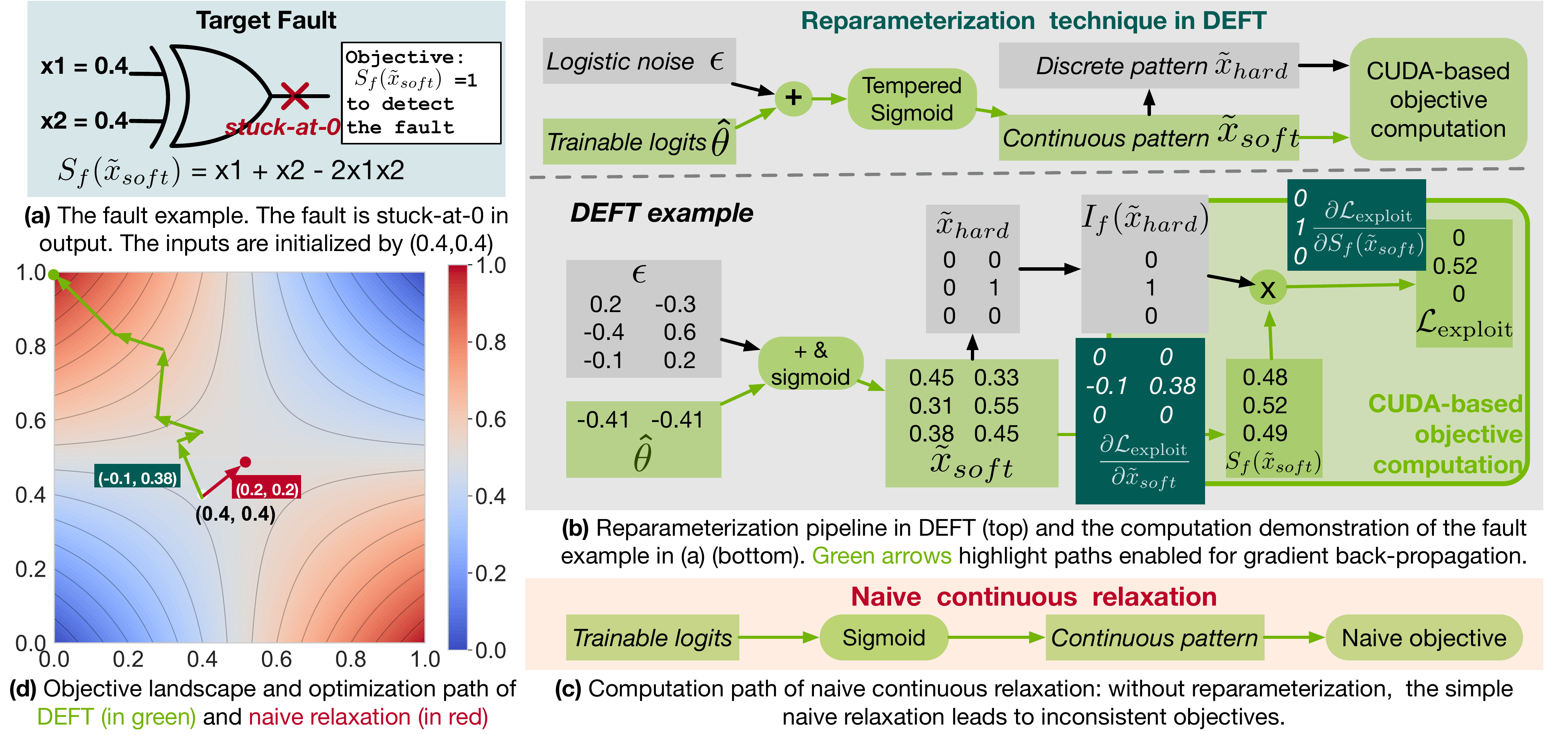}
    \caption{
        The reparameterization technique in DEFT vs. naive continuous relaxation. 
        % Naive relaxation (red arrow) optimizes continuous inputs to (0.5, 0.5) that discretize to (0,0) or (1,1), 
        % neither of which detects the fault, while DEFT's reparameterization converge to a global optimum.
        % Green arrows in right flows highlight paths enabled for gradient back-propagation.
    }
    \label{fig:gumbel}
    % \vspace*{-1em}
\end{figure*}

\new{More broadly, balancing scalability and solution quality has been a persistent challenge in VLSI design and testing. 
Complementary to learning-based and analytical approaches, differentiable programming offers an attractive alternative: 
it recasts combinatorial search as continuous optimization, enabling gradient-guided exploration 
with a global objective and efficient GPU execution. This paradigm has already delivered strong results in placement~\cite{lin2019dreamplace}, routing~\cite{li2024dgr}, and timing analysis~\cite{lu2025insta}, and is now beginning to show promise for ATPG}
by reformulating discrete test generation as 
continuous optimization on a relaxed, differentiable circuit graph.
Despite its potential for a better tradeoff between scalability and quality,
there is a fundamental conceptual barrier: 
the inherent {semantic mismatch} between continuous relaxations and the discrete, 
non-smooth semantics of fault activation and error propagation. 
A continuous optimizer may converge to some values whose corresponding binary test pattern fails to detect the fault (see an example in \Cref{fig:gumbel}).

To address the barrier, we introduce a new reparameterized formulation that 
yields a continuous objective whose expected value is provably aligned with the discrete ATPG detection semantics.
This reparameterization provides the conceptual key, but turning it into a practical ATPG engine for industrial circuits still requires solving two challenges: 
1) computational efficiency, as the forward and backward propagation through the massive computation graph of a large circuit 
creates a performance bottleneck, 
and 2) extreme graph depth, an inherent property of circuits that leads to vanishing gradient issues during optimization.

We propose \textbf{DEFT} (Differentiable Automatic Test Pattern Generation), an ATPG engine designed to systematically overcome these obstacles.
DEFT converts the circuit into a differentiable computation graph, 
constructs a reparameterization framework to capture the expected discrete detection semantics, 
and optimizes input logits using gradient ascent. To ensure practical scalability and stability, 
DEFT employs gradient normalization to stabilize optimization on extremely deep graphs and implements a custom CUDA kernel,
which achieves \textbf{4x-26x speedup} and \textbf{nearly 2x memory reduction} over a standard PyTorch+DGL \cite{wang2019deep} implementation.
Evaluated on a wide range of benchmarks, DEFT \textbf{reduced the pattern count by 27.3\%} on average and by up to 75.9\%.
DEFT is also flexible for practical ATPG requirements such as X-bit-aware pattern generation: it produces patterns with {19.3\% fewer 0/1 bits} while still {detecting 35\% more faults}.
{Our contributions are summarized as follows:}
\begin{itemize}[nosep,leftmargin=*]
\item We develop a reparameterized differentiable ATPG formulation that mathematically resolves the mismatch between continuous relaxations and discrete fault semantics.
\item We propose DEFT, a scalable and effective ATPG engine that addresses the practical challenges of computation efficiency and extreme graph depth.
\item We implement a custom CUDA kernel that achieves {4x-26x speedup} and about {2x memory savings} over a PyTorch+DGL baseline.
\item We validate DEFT on HTD fault sets from academic and industrial benchmarks and compare it with a leading commercial tool, 
demonstrating 27.3\% reduced pattern count on average and producing 19.3\% fewer specified bits in partial assignment settings.
\end{itemize}

\section{Related Work}
\label{sec:prelim}

\subsection{Automatic Test Pattern Generation (ATPG)}

Structural ATPG algorithms like PODEM \cite{goel1981implicit} and FAN \cite{fujiwara1983acceleration} 
have dominated industrial applications 
due to their high efficiency and scalability. 
Their limitation becomes apparent on hard-to-detect (HTD) faults, 
where the final 1\% of untested faults can consume over 71.8\% of the total test patterns on the NVDLA MAC benchmark.
% where the search space becomes exponentially large and backtracking dominates the runtime.

Recent work leverages machine learning to improve these heuristic algorithms.
SmartATPG \cite{li2024smartatpg}, for example, employs reinforcement learning to reduce backtracks with a better decision policy.
A primary challenge for such methods is the inference overhead introduced at each decision step, 
which struggles to compete with the CPU-cycle-level efficiency of traditional heuristics. 
% While these approaches improve heuristic quality, 
% they do not change the underlying backtracking-based search, which remains the fundamental bottleneck.

SAT-based ATPG encodes the search problem into a Boolean Satisfiability (SAT) instance. 
This approach excels at resolving complex HTD faults and is highly effective for proving fault redundancy.
However, SAT-based ATPG suffers from several fundamental limitations.
First is a scalability challenge on large designs. 
Second, as noted in the introduction, is a mismatch: 
SAT is a decision framework, not an optimization one \cite{becker2014recent}.
This makes it difficult to efficiently solve for minimal pattern sets or perform multi-target ATPG, 
leading to ``pattern inflation" \cite{chao2025pastatpg}. 
Third, SAT-based ATPG is inherently a per-fault, non-multi-target solver. 
Without the extensive global compaction heuristics available in commercial tools, 
the final test set is often much larger. 
Even hybrid approaches like PastATPG \cite{chao2025pastatpg}, 
which uses partial-assignment SAT, still generate larger test sets than highly optimized commercial tools.

\subsection{Differentiable Programming for EDA}

Differentiable programming (DP) reframes discrete problems as continuous optimization tasks, enabling gradient-based search and GPU acceleration.
It has achieved strong results in several EDA domains, including placement~\cite{lin2019dreamplace}, routing~\cite{li2024dgr}, and timing analysis~\cite{lu2025insta}.
A key reason for these successes is the semantic alignment between the continuous gradient and the underlying optimization objective.
For example, in placement, gradients correspond to physical forces~\cite{lin2019dreamplace}; in timing analysis, they correspond to timing sensitivities~\cite{lu2025insta}.

\new{Applying DP to ATPG is more challenging.
Prior work has explored continuous relaxation for ATPG by constructing a neural twin of the circuit, where each standard cell is replaced by a neural-network surrogate~\cite{tan2024safety}.
While this demonstrates the potential of gradient-based test generation, it leaves two key limitations.
First, ATPG is inherently discrete: the final output must be a binary test pattern that activates a fault and propagates its effect to an observation point.
Naive continuous relaxation can therefore suffer from a semantic mismatch, where improving the relaxed objective does not necessarily produce a valid detecting pattern after discretization.
\Cref{fig:gumbel} illustrates this issue on a simple XNOR circuit: the relaxed optimizer converges to $(0.5,0.5)$, which discretizes to either $(0,0)$ or $(1,1)$, neither of which detects the target fault.
Second, modeling every standard cell with a neural-network surrogate introduces substantial computation and memory overhead during forward and backward propagation, limiting scalability to industrial-scale circuits.
}

\new{DEFT addresses both limitations.
It introduces a reparameterized formulation whose expected continuous objective is aligned with discrete fault-detection semantics.
It also directly models gate functionality using closed-form continuous relaxations and implements levelized forward--backward propagation with a custom CUDA kernel.
This design avoids per-gate neural-network inference, improves efficiency, and enables scalable gradient-based ATPG with natural support for multi-fault objectives and global pattern sharing.
}

\section{Problem Formulation}
\label{sec:problem}

A discrete input pattern $\data \in \B^n$ {detects} fault $f$ if the output of the fault-free circuit differs from the output of the faulty circuit.
We define a discrete indicator function $I_f: \B^n \to \B$ that formalizes this.
% \begin{equation}
%     I_f(\data) = \I\left[ \max_{j \in \{1..m\}} |C(\data)_j - C_f(\data)_j| > 0.5 \right] = \begin{cases} 1 & \text{if } \data \text{ detects } f \\ 0 & \text{otherwise} \end{cases}
% \end{equation}
$I_f(\data)$ returns 1 if the pattern $\data$ detects $f$ (i.e., at least one output bit differs between the fault-free and faulty circuits) and 0 otherwise.
The goal of test pattern generation is to find any pattern $\data^*$ that detects the fault. This can be expressed as a discrete optimization problem:
\begin{equation}
    \max_{\data \in \B^n} I_f(\data)
    \label{eq:p1}
    \tag{DEFT-P1}
\end{equation}
An optimal solution $\data^*$ to \eqref{eq:p1} is any pattern for which $I_f(\data^*) = 1$.

\section{Differentiable Reparameterization for ATPG}
\label{sec:reformulation}
We present the mathematical formulation that transforms the discrete problem of ATPG 
into a continuous optimization problem, keeping the fault-detection semantics intact.
The framework consists of three stages:
\begin{enumerate}[nosep,leftmargin=*]
    \item \textbf{Discrete Problem (DEFT-P1):} The original, discrete ATPG objective, as explained in \Cref{sec:problem}.
    \item \textbf{Probabilistic Reframing (DEFT-P2):} A conceptual reframing of the problem as maximizing an expected value.
    \item \textbf{Differentiable Surrogate (DEFT-P3):} A practical, differentiable objective function based on reparameterization, which can be optimized with gradient ascent.
\end{enumerate}

\subsection{Probabilistic Reframing}
The core difficulty in ATPG is that the test pattern is inherently discrete --- changing any single bit may flip a fault from detectable to undetectable.
Directly optimizing over the \{0,1\} space (\ref{eq:p1}) is therefore non-differentiable.
To enable gradient-based search, we first reinterpret a binary pattern as a probability distribution over patterns, and then optimize its expected fault-detection probability. This leads to the probabilistic formulation in DEFT-P2.
% Equation DEFT-P1 is non-differentiable because of discrete inputs,
% we reframe this problem by parameterizing a probability distribution over the discrete space $\B^n$.

Let $\param \in [0, 1]^n$ be a vector of continuous parameters, where each $\param_i$ defines the probability for an independent Bernoulli distribution. This defines a product distribution $\prob$ over $\B^n$:
\begin{equation}
    p(\data | \param) = \prod_{i=1}^n \param_i^{\data_i} (1 - \param_i)^{1 - \data_i}
\end{equation}
We can now reframe the optimization goal: instead of finding a single discrete pattern $\data$, we seek optimal parameters $\param^*$ that maximize the \textit{expected} detection probability over the distribution $\prob$.
\begin{equation}
    \max_{\param \in [0, 1]^n} J_f(\param) \quad \text{where} \quad J_f(\param) = \E_{\data \sim \prob} [I_f(\data)]
    \label{eq:p2}
    \tag{DEFT-P2}
\end{equation}

Then the following statement holds; its proof is given in Appendix~\ref{app:proof-equivalence}.
\begin{theorem}[Formulation Equivalence]
    \label{thm:equivalence}
    The discrete problem \eqref{eq:p1} from \Cref{sec:problem} has a solution (i.e., a detecting pattern $\data^*$ exists) if and only if the optimal value of the probabilistic problem \eqref{eq:p2} is 1.
    \begin{equation*}
        \max_{\data \in \B^n} I_f(\data) = 1 \quad \iff \quad \max_{\param \in [0, 1]^n} J_f(\param) = 1
    \end{equation*}
    \end{theorem}

While maximizing this expectation in DEFT-P2 is tautologically equivalent to finding a distribution whose probability mass is 
concentrated on detecting patterns (i.e., $J(\param^*) = 1$, which implies all non-detecting patterns have zero probability), 
this formulation is not directly solvable. 
Its value is conceptual: it shifts the problem from a discrete search for $\data$ to a continuous optimization of the parameters $\param$.
 The core technical challenge is to formulate a differentiable estimator for this expectation.

\subsection{Differentiable Surrogate via Reparameterization}

To optimize \eqref{eq:p2} with gradients, we must overcome the non-differentiable sampling operation $\data \sim \prob$. 
We use the Gumbel-Softmax reparameterization trick \cite{jang2016categorical} to create a differentiable computation path, including the following steps:

\begin{enumerate}[leftmargin=*]
    \item \textbf{Logit-Space Parameterization:} We reparameterize $\param \in [0, 1]^n$ with unconstrained logits 
    $\logits \in \R^n$, where $\param_i = \sigma(\logits_i)$.
    
    \item \textbf{Reparameterization:} We rewrite the discrete sample $\data$ as a deterministic function of the 
    logits $\logits$ and an independent Logistic noise vector $\noise$. 
    % We use Logistic noise, as seen in the DEFT implementation and the DiffILO paper.
    Define the discrete pattern $\xh$ as: 
    \begin{equation}
        \xh(\logits, \noise)_i = \I(\logits_i + \noise_i > 0)
    \end{equation}
% \begin{lemma}[Logistic Reparameterization]
%     Let $\logits \in \R^n$ be a vector of logits. Let $\noise \in \R^n$ be a noise vector with independent components $\noise_i \sim \text{Logistic}(0, 1)$, drawn by sampling $u_i \sim \mathcal{U}(0, 1)$ and setting $\noise_i = \log(u_i) - \log(1 - u_i)$.
%     Let the discrete pattern $\xh$ be defined as:
%     \begin{equation}
%         \xh(\logits, \noise)_i = \I(\logits_i + \noise_i > 0)
%     \end{equation}
%     Then the distribution of $\xh$ is identical to $\prob$ where $\param_i = \sigma(\logits_i)$. That is, $P(\xh(\logits, \noise)_i = 1) = \sigma(\logits_i)$.
%     \end{lemma}
    Using logistic reparameterization (proved in Appendix~\ref{app:proof-reparam}), 
    the expectation from \eqref{eq:p2} can be rewritten as:
    \begin{equation}
        J_f(\logits) = \E_{\noise \sim \text{Logistic}(0,1)^n} [ I_f(\xh(\logits, \noise)) ]
        \label{eq:p2_logits}
    \end{equation}
    
    \item \textbf{Hybrid Surrogate Objective:} The objective in Equation \ref{eq:p2_logits} is still not differentiable:
    The term $I_f(\xh)$ is a 0/1 indicator, and its gradient with respect to $\logits$ is zero almost everywhere.
    To create a trainable signal, we must replace the discrete objective $I_f(\xh)$ inside the expectation with a differentiable surrogate.
    Inspired by the method used in \cite{geng2025differentiable}, we create a hybrid surrogate that combines the discrete indicator with a continuous, differentiable signal.

    \begin{itemize}
        \item A \textbf{continuous surrogate pattern} $\xs$, which is the Gumbel-Softmax relaxation:
        \begin{equation}
            (\xs)_i(\logits, \noise, \tau) = \sigma\left( \frac{\logits_i + \noise_i}{\tau} \right)
        \end{equation}
        where $\tau$ is an annealed temperature parameter.
        
        \item A \textbf{differentiable surrogate signal}  $S_f: [0, 1]^n \to \R^+$, which is the continuous relaxation of the detection function $I_f$ and fully differentiable w.r.t. $\logits$.
        % $S_f(\xs)$ is designed to be positive and large when $\xs$ ``looks like" a detecting pattern, and it is 
    \end{itemize}

We now formally define our new objective, $\mathcal{L}^f_{\text{exploit}}$, 
by replacing the discrete $I_f(\xh)$ in Equation \ref{eq:p2_logits} with this hybrid surrogate, $S_f(\xs) \cdot I_f(\xh)$.
% We combine these into the \textbf{``exploit" objective} $\mathcal{L}^f_{\text{exploit}}$. 
% This loss function uses the discrete pattern $\xh$ to ``gate" the gradient from the continuous pattern $\xs$:
\begin{equation}
    \mathcal{L}^f_{\text{exploit}}(\logits) = \E_{\noise} \left[ \underbrace{S_f(\xs(\logits, \noise, \tau))}_{\text{Continuous Surrogate}} \cdot \underbrace{I_f(\xh(\logits, \noise))}_{\text{Discrete Indicator}} \right]
    \label{eq:p3}
    \tag{DEFT-P3}
\end{equation}

This formulation is the core of the ``exploit" objective. It includes two key components: The \textbf{discrete indicator} $I_f(\xh)$ acts as a ``gate." It ensures the objective is zero if the discrete pattern fails to detect the fault. The \textbf{continuous surrogate} $S_f(\xs)$ provides a non-zero, differentiable signal (a gradient path) when the discrete gate is ``open" (i.e., $I_f(\xh) = 1$).

By maximizing $\mathcal{L}^f_{\text{exploit}}$, we optimize the logits $\logits$ to produce samples that are both discretely correct ($I_f=1$) and have a strong continuous signal ($S_f > 0$).
\end{enumerate}
\subsection{Gradient Calculation of the Exploit Signal}

The gradient of the ``exploit" objective \eqref{eq:p3} is computed by treating the discrete indicator $I_f(\xh)$ as a constant with respect to the gradient operator. Since the gradient of the discrete indicator $I_f(\xh)$ is zero almost everywhere, we adopt a straight-through (ST) estimator that ignores the gradient through $I_f(\xh)$ and only propagates gradients through the continuous surrogate path $S_f(\xs)$.
\begin{align}
    \nabla_{\logits} \mathcal{L}^f_{\text{exploit}}(\logits) &= \nabla_{\logits} \E_{\noise} \left[ S_f(\xs) \cdot I_f(\xh) \right] \\
    % &= \E_{\noise} \left[ \nabla_{\logits} \left( S_f(\xs(\logits, \noise, \tau)) \cdot I_f(\xh(\logits, \noise)) \right) \right] \\
    &\approx \E_{\noise} \left[ \underbrace{\nabla_{\logits} S_f(\xs(\logits, \noise, \tau))}_{\text{Continuous gradient path}} \cdot \underbrace{I_f(\xh(\logits, \noise))}_{\text{Discrete indicator}} \right] \label{eq:grad_exploit_exact}
\end{align}
This expectation is estimated using Monte Carlo sampling, i.e., by averaging over $K$ noise samples:
\begin{equation}
    \nabla_{\logits} \mathcal{L}^f_{\text{exploit}}(\logits) \approx \frac{1}{K} \sum_{k=1}^K \left[ \nabla_{\logits} S_f(\xs^{(k)}) \cdot I_f(\xh^{(k)}) \right]
    \label{eq:grad_exploit}
\end{equation}

\subsection{Handling Zero Gradients: The Explore Objective}
A problem arises with the gradient in \Cref{eq:grad_exploit}.
In the early training stage, the probability $P(I_{f}(\tilde{x}_{hard})=1)$ is often extremely low, particularly for complex and deep circuits.
This means it is highly likely that no sampled pattern $\tilde{x}_{hard}^{(k)}$ will detect the fault, resulting in $I_{f}(\tilde{x}_{hard}^{(k)})=0$ for all $K$ samples,
causing the entire gradient estimate to collapse to zero and leading to optimization stagnation.
To solve this, we introduce a complementary \textbf{``explore" objective}, $\mathcal{L}^f_{\text{explore}}$,
serving as a differentiable proxy for the discrete goal and a bootstrap mechanism: 
when $I_{f}(\tilde{x}_{hard})=0$ silences the exploit gradient, $\mathcal{L}_{\text{explore}}^{f}$ ensures a continuous, non-zero differentiable signal is maintained.
$\mathcal{L}^f_{\text{explore}}(\logits)$ is defined as the expected continuous surrogate signal $\mathcal{L}^f_{\text{explore}}(\logits) = \E_{\noise} \left[ S_f(\xs(\logits, \noise, \tau)) \right]$.
% \begin{equation}
%     \mathcal{L}^f_{\text{explore}}(\logits) = \E_{\noise} \left[ S_f(\xs(\logits, \noise, \tau)) \right]
% \end{equation}
% This encourages the optimizer to maximize the continuous fault-detection signal, which serves as a differentiable proxy for the discrete goal, 
% guaranteeing a non-zero gradient to steer $\hat{\theta}$
The final objective combines these terms using an annealing schedule $w_{\text{explore}}$ to dynamically control the training focus:
\begin{equation}
\label{eq:deft:final_obj}
    \mathcal{L}^f_{\text{final}}(\logits) = (1 - w_{\text{explore}}) \cdot \mathcal{L}^f_{\text{exploit}}(\logits) + w_{\text{explore}} \cdot \mathcal{L}^f_{\text{explore}}(\logits)
\end{equation}
$w_{\text{explore}}$ transitions from 1 (prioritizing rapid search in the continuous active region) to 0
(prioritizing the detection of a discretely valid pattern). 
This transition is crucial because only $\mathcal{L}_{\text{exploit}}^{f}$ is mathematically aligned with the true discrete fault-detection semantics, 
guaranteeing a discretely correct final pattern.

\subsection{Intuition behind the formulation}
\new{The idea of interpreting logic signals as probabilities, and propagating these values through Boolean networks,
has been studied since early work on probabilistic circuit analysis and test generation decades ago
\cite{parker1975probabilistic,parker1975analysis,mccluskey1978boolean}. 
In recent years, similar continuous
logic relaxations have reappeared in differentiable optimization for combinatorial problems, including logic
synthesis \cite{wang2024towards} and neural architecture search \cite{petersen2022deep}. 
% This view is attractive
% for DEFT because it converts a discrete circuit into a smooth computation graph. 
However, classical probabilistic
logic propagation is exact only under the independence assumptions of input signals \cite{parker1975probabilistic}. 
When fanout
branches reconverge, the same upstream random variable can reach multiple inputs of a downstream gate; treating
these inputs as independent introduces a correlation error.
}

\new{Prior work already recognized this issue for reconvergent fanout \cite{parker1975probabilistic, parker1975analysis}. 
Exact or more faithful treatments can be
obtained by symbolic probability manipulation, such as suppressing exponents in probability expressions
\cite{parker1975probabilistic}, or by representing Boolean functions with ordered binary decision diagrams
\cite{bryant1986graph}. These approaches are valuable for exact analysis, but not a good fit for DEFT's
goal: scalable gradient optimization over large circuits using simple, levelized GPU kernels. 
DEFT therefore keeps the local differentiable logic surrogate, but adds two safeguards:
(1) The hard indicator $I_f$ to ensure the error from reconvergence does not lead to false optimization directions, and
(2) Monte Carlo sampling over $\epsilon$ partially preserves the variance and correlation terms induced by reconvergence, thereby reducing the approximation error. 
The following examples give intuition
for these two roles.
}

\begin{figure}
    \centering
    \includegraphics[width=0.4\textwidth]{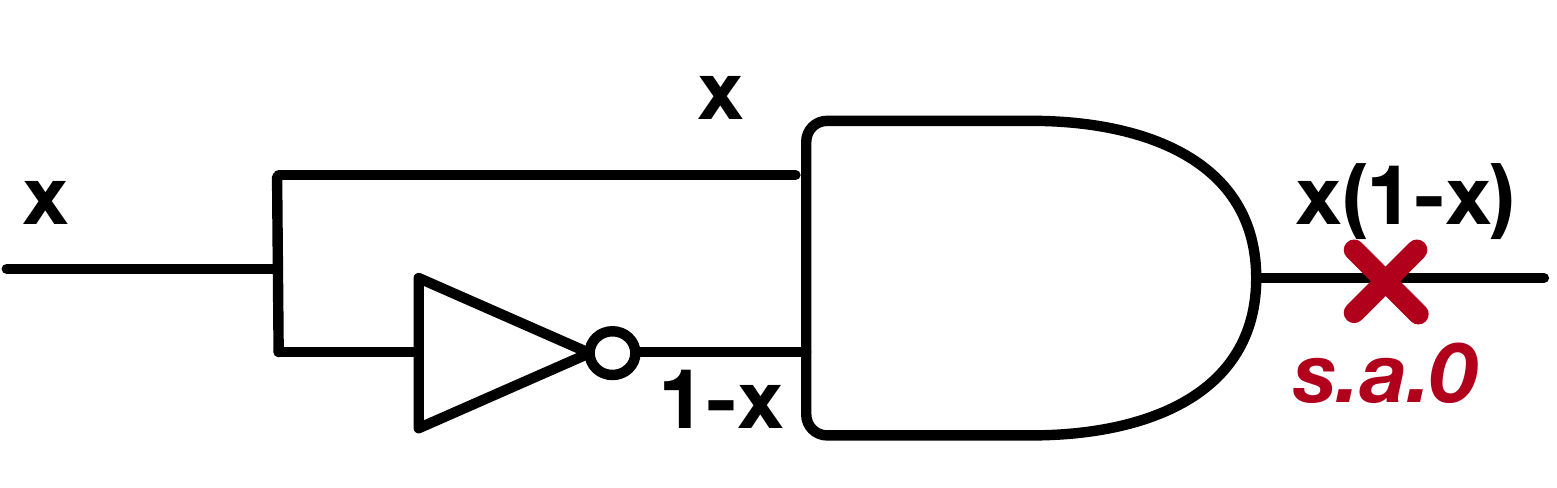}
    \caption{\new{A reconvergent example where the naive relaxation gives a non-zero gradient $1-2x$ for an
    untestable fault. Including the discrete indicator $I_f$ in \Cref{eq:grad_exploit_exact} correctly gates the
    exploit gradient to zero because no hard pattern detects the fault.}}
    \label{fig:reconvergence}
\end{figure}

\subsubsection{The Role of the Discrete Indicator \texorpdfstring{$I_f$}{If}}

\new{\Cref{fig:reconvergence} illustrates why the hard indicator is needed. In this circuit, the target fault is
untestable in the Boolean domain.
% for every binary input assignment, the fault-free and faulty outputs are
% identical. 
A purely continuous relaxation can still produce a non-zero surrogate gradient for this untestable fault, 
which may affect the detection of other faults in a multi-fault setting.
DEFT avoids this failure mode by multiplying the continuous signal by the hard detection indicator in
\Cref{eq:p3}. During optimization, a sample contributes gradient only when the corresponding hard pattern
\(\xh\) detects the fault. For an untestable fault, \(I_f(\xh)=0\) for every sample, so the exploit gradient is zero
everywhere. 
% This behavior is intentional: it prevents DEFT from converting a fractional reconvergence artifact into
% a false detection claim. 
% The explore term may still provide a smooth search signal early in training, but the
% annealing schedule eventually gives control to the exploit objective, and final patterns are always checked in the
% discrete circuit.
}

\subsubsection{The Role of Monte Carlo Sampling over \texorpdfstring{$\epsilon$}{epsilon}}
\begin{figure}
    \centering
    \includegraphics[width=0.35\textwidth]{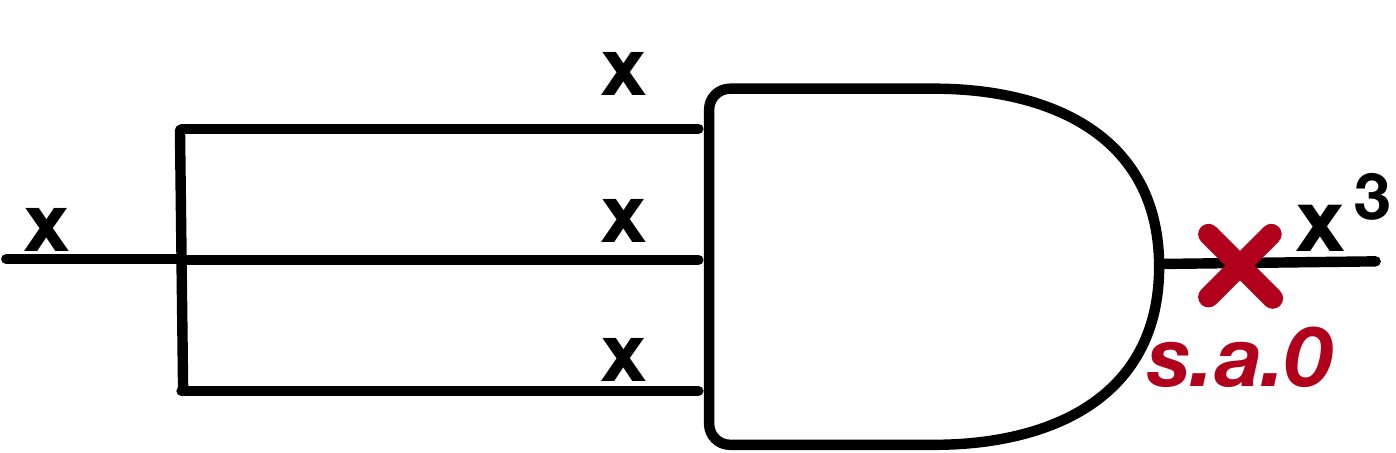}
    \caption{\new{A reconvergent structure where deterministic mean propagation yields local gradient
    $3\bar{x}^2$, while Monte Carlo sampling preserves the additional variance term
    $3\mathrm{Var}_{\noise}(x(\noise)) > 0$.}}
    \label{fig:whyrandom}
\end{figure}

\new{The second issue is not about false detection, but gradient bias. 
The bias arises because products of correlated intermediate signals generally make
\(\mathbb{E}[S(\mathbf{x})]\) different from \(S(\mathbb{E}[\mathbf{x}])\) (Jensen's Inequality).
Consider the reconvergent structure in
\Cref{fig:whyrandom}, abstracted as \(Y=X^3\), where the same upstream signal fans out and reaches three inputs of
a downstream gate. Let \(x(\noise)=\sigma((\theta+\noise)/\tau)\) be the soft value produced by one
reparameterization-noise sample. A deterministic relaxation first collapses the signal to its mean
\(\bar{x}=\mathbb{E}_{\noise}[x(\noise)]\) (in other words, its probability/expectation of signal value being one), and evaluates
\(L_{\mathrm{naive}}=\bar{x}^3\). Its local derivative with respect to the relaxed signal is
\(g_{\mathrm{naive}}=3\bar{x}^2\).}

\new{Monte Carlo estimation instead evaluates the reconvergent structure under each sampled value:
\(L_{\mathrm{MC}}=\mathbb{E}_{\noise}[x(\noise)^3]\). At the local gate level, the corresponding derivative is
\[
g_{\mathrm{MC}}=\mathbb{E}_{\noise}[3x(\noise)^2]
=3\bar{x}^2+3\mathrm{Var}_{\noise}(x(\noise)).
\]
The additional variance term is precisely the information lost when the reconvergent signal is replaced by a single
mean value before propagation. Therefore, sampling over \(\noise\) does not make the surrogate exact, but it reduces
the independence error by evaluating correlated branches consistently within each sample. As \(\tau\) decreases,
the sampled soft values become closer to binary values, making this sampled propagation better aligned with the
discrete behavior that ATPG ultimately requires.}

% The gradient of this final loss is then a weighted sum of the explore and exploit gradients:
% \begin{equation}
%     \label{eq:final_grad}
%     \nabla_{\logits} \mathcal{L}^f_{\text{final}}(\logits) \approx \frac{1}{K} \sum_{k=1}^K \left[ \nabla_{\logits} S_f(\xs^{(k)}) \cdot \left( (1 - w_{\text{explore}}) \cdot I_f(\xh^{(k)}) + w_{\text{explore}} \right) \right]
% \end{equation}
% It robustly directs the optimization of $\logits$ by initially exploring for promising signals ($w_{\text{explore}} \approx 1$) and later exploiting those signals to find true discrete detections ($w_{\text{explore}} \approx 0$).

\section{DEFT Framework}
\label{sec:framework}
In this section, we present the DEFT framework.
\Cref{sec:deft:flow} outlines the overall workflow,
\Cref{sec:deft:scale} describes the scalability mechanisms, namely the custom CUDA kernel and gradient normalization, and
\Cref{sec:deft:extension} introduces the joint multi-fault and multi-pattern formulation and extensions such as partial assignment ATPG.

\subsection{DEFT Framework and Graph Construction}
\label{sec:deft:flow}
\begin{figure}[h]
    \centering
    \hspace*{-0.2cm} 
    \includegraphics[width=.5\textwidth]{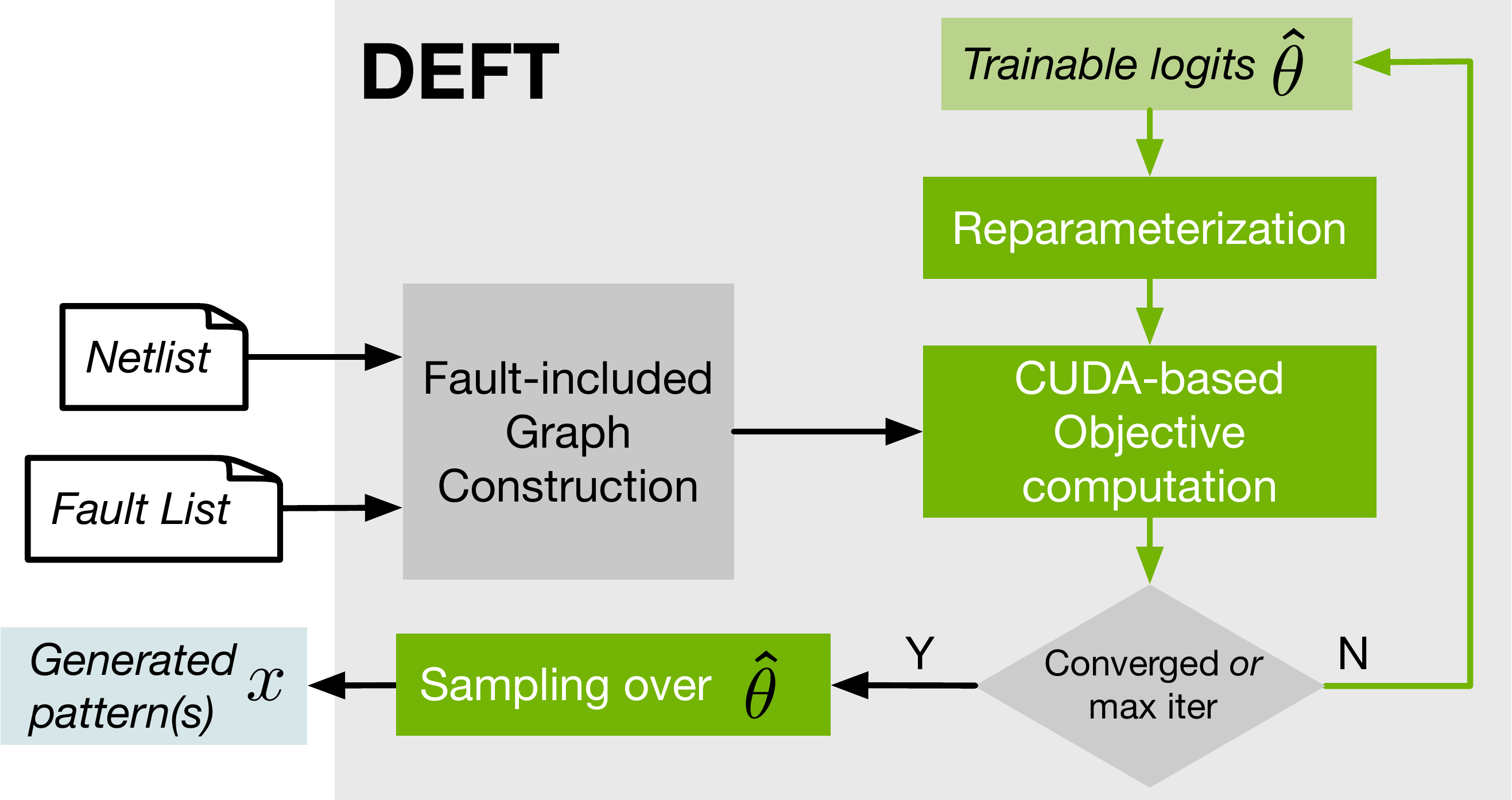}
    \caption{Overview of the DEFT Framework.}
    \label{fig:deft:flow}
    % \vspace{-0.3cm}
    \end{figure}

\Cref{fig:deft:flow} illustrates the workflow of DEFT. 
Given a circuit netlist and a target fault list $\gF$, 
we first transform the circuit into a differentiable computation graph.
We follow the well-established approach \cite{petersen2022deep, geng2025differentiable} 
by modeling the circuit as a directed acyclic graph (DAG) whose nodes are logic gates and edges are wires. 
For nets with fan-out, we introduce a virtual node to explicitly represent each branch.
The Boolean function of each gate is replaced with a smooth relaxation over $[0,1]$ so that the gradients can flow end-to-end. 
In this work, we use simple probabilistic logic as a surrogate. 
\Cref{tab:prob_model} summarizes the probabilistic functions used for general 2-input gates, 
and DEFT supports arbitrary gate types by defining their corresponding continuous relaxations.

\setlength{\abovecaptionskip}{4pt}
    
\begin{figure}
    \centering
    
    \includegraphics[width=.5\textwidth]{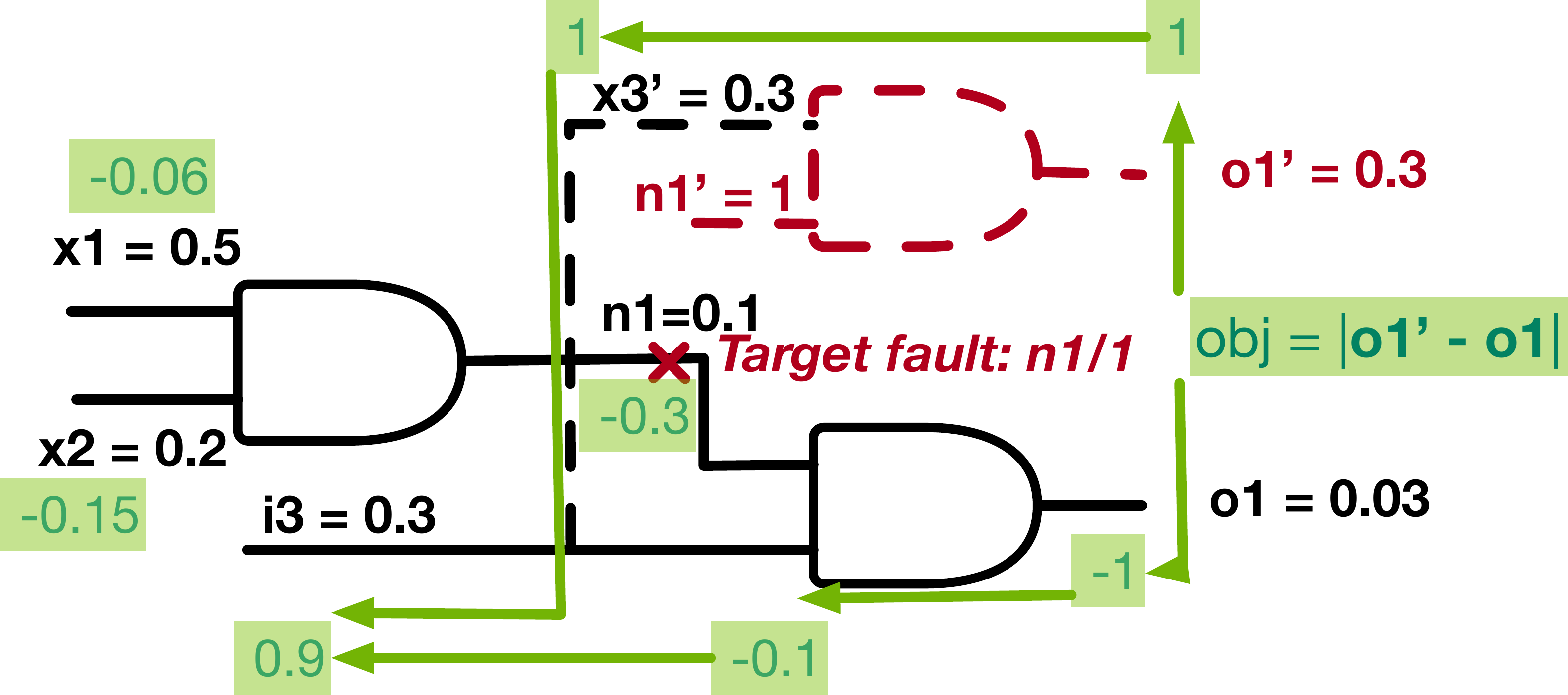}
    \caption{An example of DEFT for a stuck-at-1 fault at net n1. 
    \textcolor{nvidia}{Green} represents gradient flow, and \textcolor{myred}{red} denotes the copied fanout cone introduced by the fault.}
    \label{fig:deft:atpg-example}
    % \vspace*{-0.2cm}
\end{figure}

For each target fault $f \in \gF$, DEFT constructs a \emph{fault-included graph}. 
This involves duplicating the fan-out cone of the fault site (or the union of cones for multi-site faults) 
into fault-free (the original) and faulty copies. An example is shown in \Cref{fig:deft:atpg-example}.

\begin{table}[t]
  \hspace*{-0.3cm}
\resizebox{.5\textwidth}{!}{  \begin{tabular}{c|c|c|c|c|c|c}
      \toprule
     Gate& AND & NAND & OR     & NOR        &  XOR     & XNOR      \\
    Function & $ab$  & $1-ab$ & $a+b-ab$ & $(1-a)(1-b)$ &  $a+b-2ab$ & $1-a-b+2ab$ \\ \bottomrule
  \end{tabular}}
  \caption{Examples of 2-input ($a,b$) gate and corresponding continuous functions used in DEFT.}
  \label{tab:prob_model}
  \vspace{-.2cm}
\end{table}

\begin{figure}[h]
    \centering
    \hspace*{-0.6cm}
    \includegraphics[width=.48\textwidth]{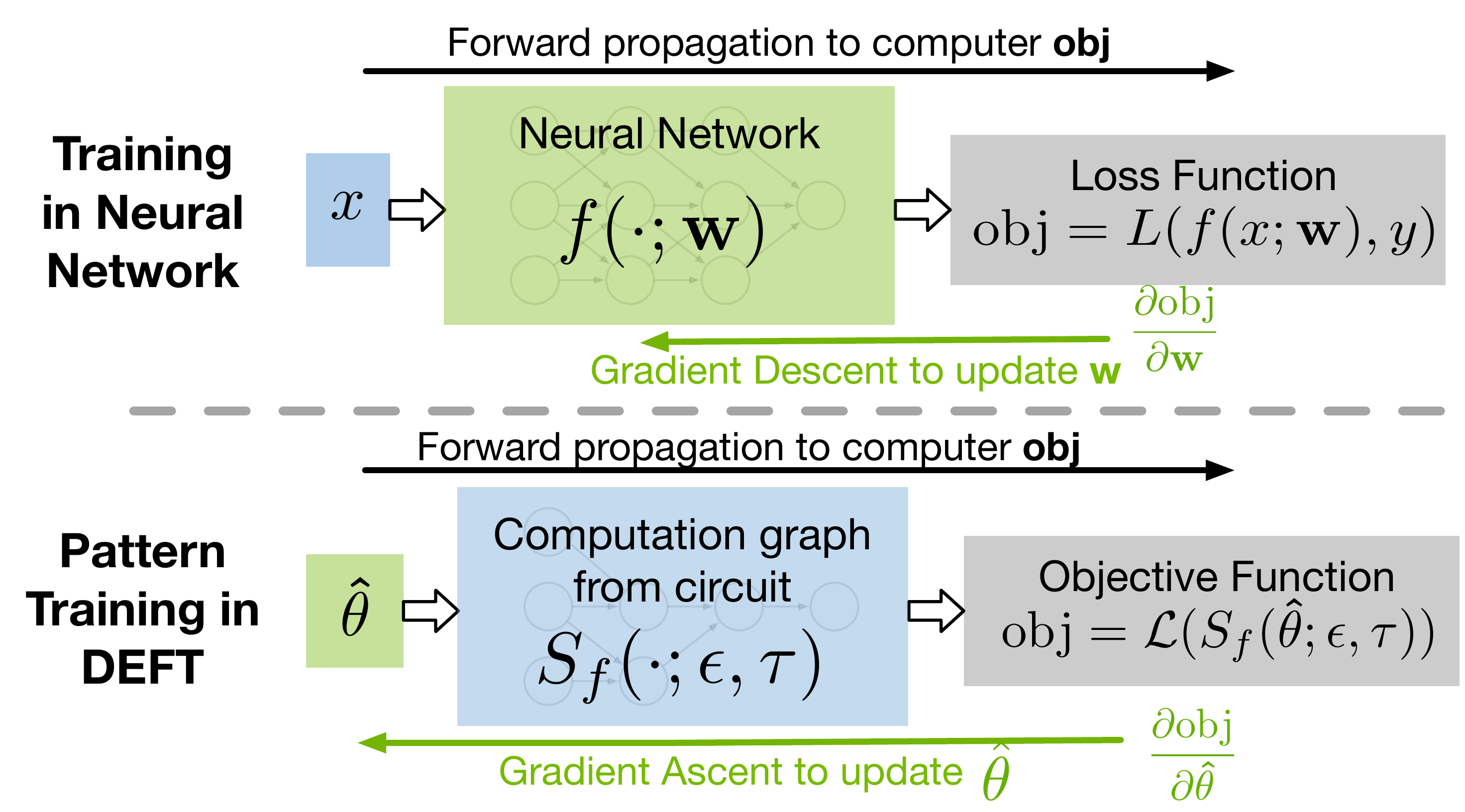}
    \caption{Analogy Between Neural Network Training and Differentiable ATPG in DEFT. 
    \textcolor{nvidia}{Green} (\textcolor{myblue}{blue}) boxes represent \textcolor{nvidia}{trainable} (\textcolor{myblue}{fixed}) components.}
    \label{fig:deft:nn}
    \end{figure}

The core of DEFT is an optimization loop that optimizes a set of trainable input logits $\logits \in \R^{n}$, 
which parameterize the relaxed input space. \Cref{fig:deft:nn} highlights the analogy between this process and 
standard neural network training. In each iteration, these logits are passed through the reparameterization 
step (detailed in \Cref{sec:reformulation}) to generate $K$ continuous samples $\xs$.
These samples are used to compute the continuous surrogate signal $S_f(\xs)$, 
which is the relaxed detection function:
\begin{equation}
  \label{eq:s_f}
S_f(\xs) = \max_{j \in {\text{POs}}} | \gG(\xs)_{{PO}_j}^{\text{fault-free}} - \gG(\xs)_{{PO}_j}^{\text{faulty}} |
\end{equation}
where $\gG(\xs)_{{PO}_j}^{\text{fault-free}}$ 
and $\gG(\xs)_{{PO}_j}^{\text{faulty}}$ are the $j$-th PO values from their respective fan-out cones, evaluated through the continuous computation graph.
Note that while \Cref{eq:s_f} is for stuck-at-fault, 
DEFT can be readily extended to support general fault tuples \cite{blanton2006defect}. 
For instance, the objective can be formulated to satisfy condition tuples by minimizing $|\gG(\xs)_i - \text{\texttt{condition\_value}}_i|$.
We leave the exploration of other advanced fault models for DEFT as future work.

Combining \Cref{eq:s_f} and \Cref{eq:deft:final_obj}, gradients of the final objective are back-propagated to 
update $\logits$ via gradient 
ascent. 
\Cref{fig:deft:atpg-example} gives an example of value propagation and gradient backpropagation.
After convergence or a maximum iteration limit, DEFT samples from the optimized logits $\hat{\theta}$ to produce the final discrete test pattern.

% \subsection{Digital Circuit to Differentiable Computation Graph}
% \label{sec:alg}

% By systematically replacing all gates, the entire circuit becomes a deep computation graph implementing a function $Y = F(X)$, 
% where $X$ and $Y$ are continuous-valued vectors. This transformation allows us to use automatic differentiation to efficiently compute the gradient of any output with respect to any input, $\nabla_{X} Y$. The levelized, feed-forward nature of combinational logic is ideally suited for the forward and backward passes of automatic differentiation, enabling massive GPU parallelization. This differentiable graph serves as the core engine for both forward simulation (evaluating a pattern) and backward propagation (computing optimization gradients).

% \begin{figure}
%     \centering
%     \includegraphics[width=.4\textwidth]{figs/example_FRI.pdf}
%     \caption{An example of continuous relaxations for standard logic gates.}
%     \label{fig:deft:fri}
% \end{figure}

\subsection{Scaling to Industrial Circuits}
\label{sec:deft:scale}
Scaling DEFT to industrial circuits requires solving two practical challenges: 
computational efficiency and optimization stability for extremely deep graphs. 
To address these issues, we present two key techniques in this section: a custom CUDA kernel and gradient normalization.
\subsubsection{Custom CUDA Kernel for Efficient Propagation}
\label{sec:cuda}
\begin{figure}
  \centering
  \includegraphics[width=.5\textwidth]{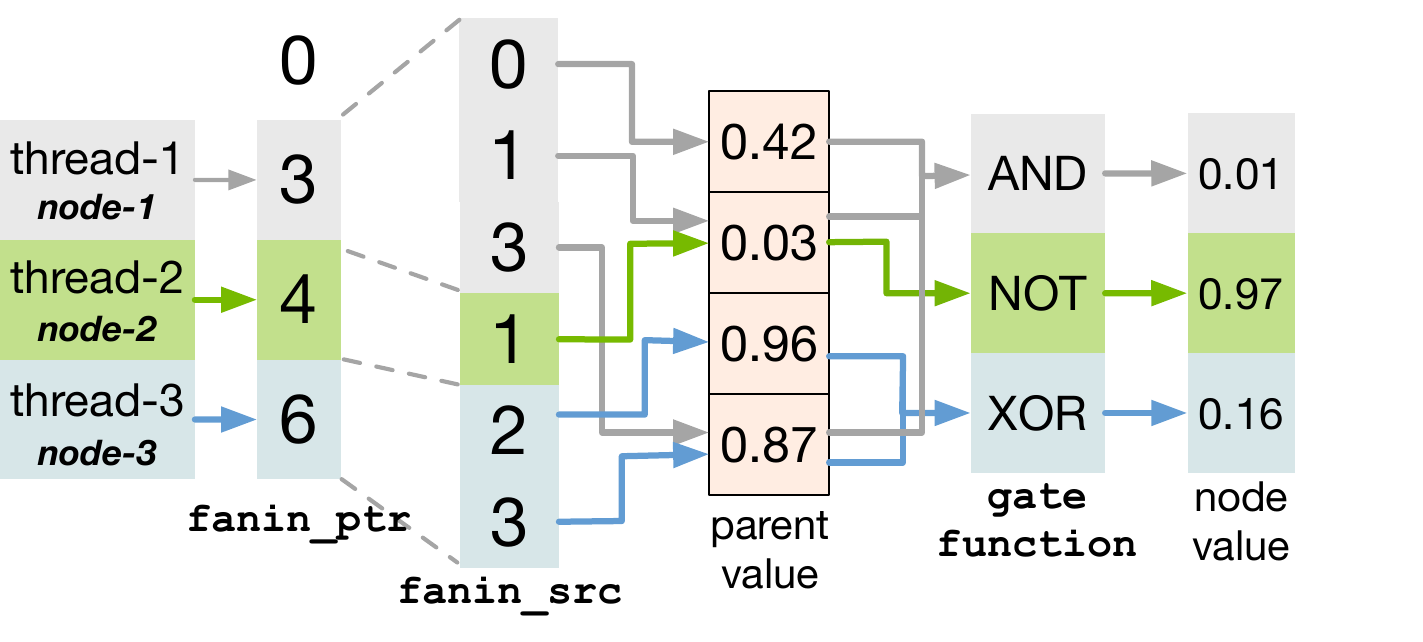}
  \caption{Illustration of continuous value propagation using custom CUDA kernel in DEFT.
  }
  \label{fig:deft:cuda}
  % \vspace*{-0.1cm}
\end{figure}

Our simulator exploits GPU parallelism to propagate values over the large computation graphs. 
At each topological level, the host identifies the nodes in that level and launches a CUDA kernel. 
As shown in \Cref{fig:deft:cuda}, an index array stored on the GPU maps each CUDA thread to a gate. 
Using a node-centric Compressed Sparse Row structure (CSR) layout (\texttt{fanin\_ptr}, \texttt{fanin\_src}), each thread locates its fan-in range, 
gathers parent value, computes the output based on the gate function, and writes the result to a level-output tensor, which is then scattered back to the global node tensor. 
This design replaces DGL's generic \texttt{pull} abstraction with a single fused kernel that performs fan-in reduction directly on 
dense tensors with coalesced memory accesses.

For end-to-end training, the kernel is exposed to PyTorch as a custom autograd operator. 
During the forward pass, the C++ wrapper stores only the information required for backpropagation—
the level indices, CSR structures, gate functions, and a compact buffer of fan-in values—while omitting global node values, 
substantially reducing memory usage compared to a naive autograd trace over all primitive operations.

\subsubsection{Gradient Normalization for Deep Graphs}
% The second scaling challenge is optimization stability.
In probabilistic-logic relaxations, per-gate derivatives are bounded within $[-1,1]$. 
Composed across an inherently deep computation graph, 
this leads to vanishing gradients and slow training. 
As a pragmatic stabilization, we apply $\ell_2$ normalization for all gradients w.r.t. input logits $\blogits$. 
We note that while effective, this is a stabilization technique rather than a fundamental solution. 
A promising direction for future work is the design of continuous relaxations for logic gates that are less susceptible to vanishing gradients.

\subsection{Multi-Target Optimization and Extensions}
\label{sec:deft:extension}
% In this section, we introduce several key techniques that enhance the effectiveness and efficiency of DEFT.
\subsubsection{Joint Multi-Fault and Multi-Pattern Optimization}
Unlike traditional ATPG methods that treat faults independently and rely on post-processing for compaction, 
DEFT naturally supports joint multi-fault and multi-pattern optimization. 
By merging the fan-out cones of all target faults $\gF$ into a single computational graph, 
DEFT optimizes a batch of $T$ candidate patterns, 
parameterized by logits $\blogits \in \mathbb{R}^{n \times T}$, against a unified objective:

\begin{equation}
    \mathcal{L}_{\text{multi}}(\blogits) = \sum_{f \in \gF} \max_{t \in \{1,\ldots,T\}} \left(\mathcal{L}^f_{\text{final}}(\logits_t) \right) 
\end{equation}

% [Comment: should we mention the benefits of multi-pattern optimization here? e.g. test pattern co-optimize, so better quality?]

Moreover, DEFT naturally extends to a batched run by optimizing $B$ independently initialized pattern sets in parallel, 
yielding $\blogits\in\mathbb{R}^{n\times T\times B}$, which stabilizes convergence.

\subsubsection{Extensions to Practical ATPG Requirements}
\label{sec:deft:extensions}
DEFT's differentiable objective readily accommodates common industrial constraints.
For example, to produce patterns with more don't-cares (X-bits), which facilitates downstream compaction, we add an X-promoting penalty:
\begin{equation}
  \label{eq:deft:x_penalty}
  \mathcal{L}_{\text{X}}(\blogits) = -\lambda \sum_{i=1}^{n} \sum_{t=1}^{T} |{(\xs)}_i^t - 0.5|,
\end{equation}
where $\lambda$ controls the strength. Under gradient ascent, this term penalizes unnecessary distance from 0.5 and pushes logits toward ambiguity when the bit is not critical for detecting any target fault.
% Similarly, for low-power scan-shift ATPG, a regularization term 
% can minimize the Hamming distance between consecutive patterns: $\sum_{t=2}^{T} \sum_{i=1}^{n} |(\xs)^t - (\xs)^{t-1}|$. A similar L1 distance penalty can be applied for low-power scan-capture ATPG by minimizing transitions between pre-capture and post-capture states.

This flexibility extends to other requirements, such as low-power ATPG, 
up-weight detection on safety-critical POs (e.g., brake system in an automotive design),
or prioritizing faults on timing-critical paths.
% Because all penalties are differentiable, 
DEFT can co-optimize the primary detection objective with these practical constraints in a single end-to-end framework.

\section{Experiments}
\label{sec:result}

\begin{figure}
    \centering
    \includegraphics[width=0.45\textwidth]{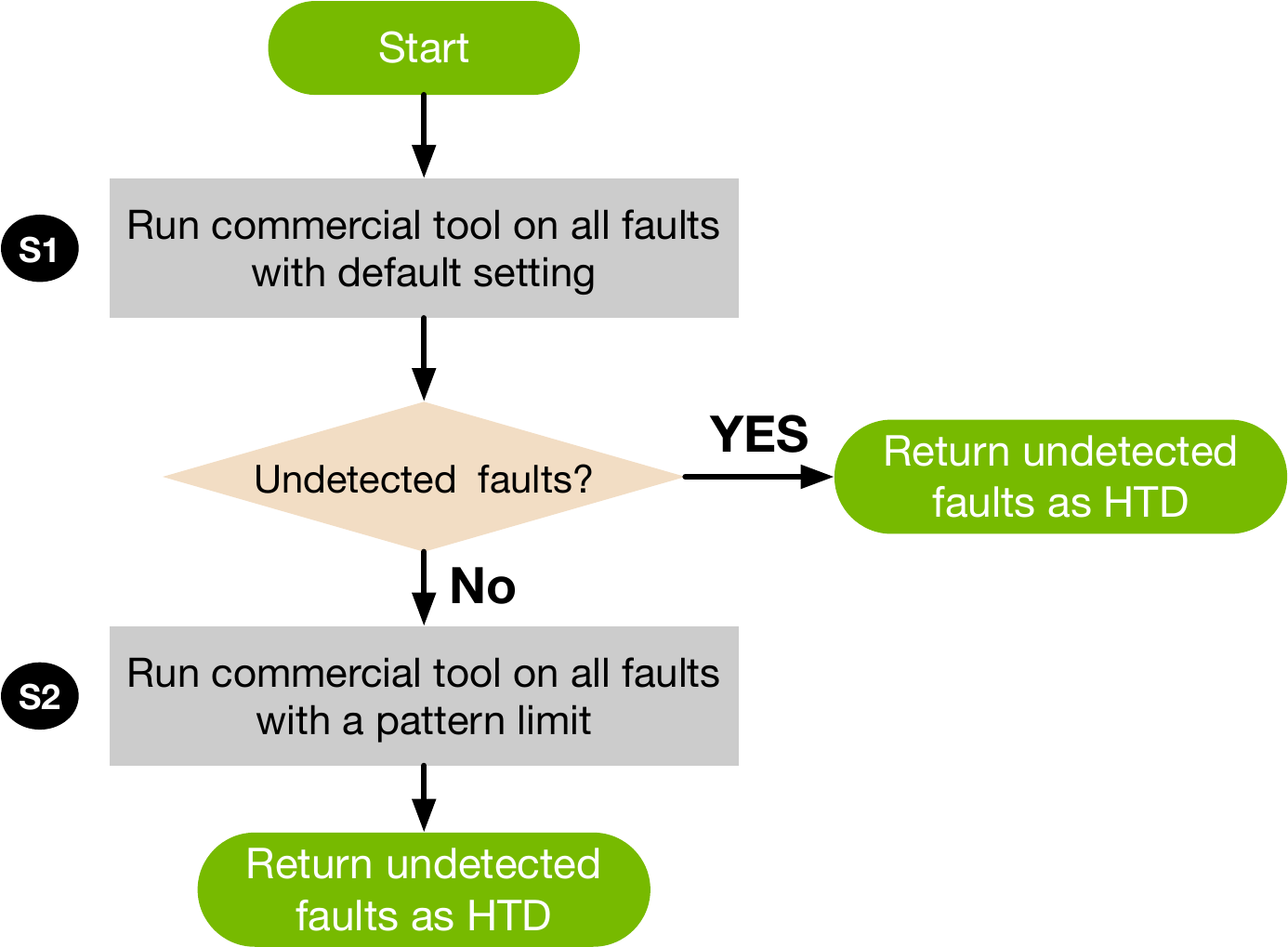}
    \caption{The process of identifying hard-to-detect (HTD) faults using a commercial ATPG tool.}
    \label{fig:htd_process}
\end{figure}

\begin{figure}
    \centering
    \includegraphics[width=0.5\textwidth]{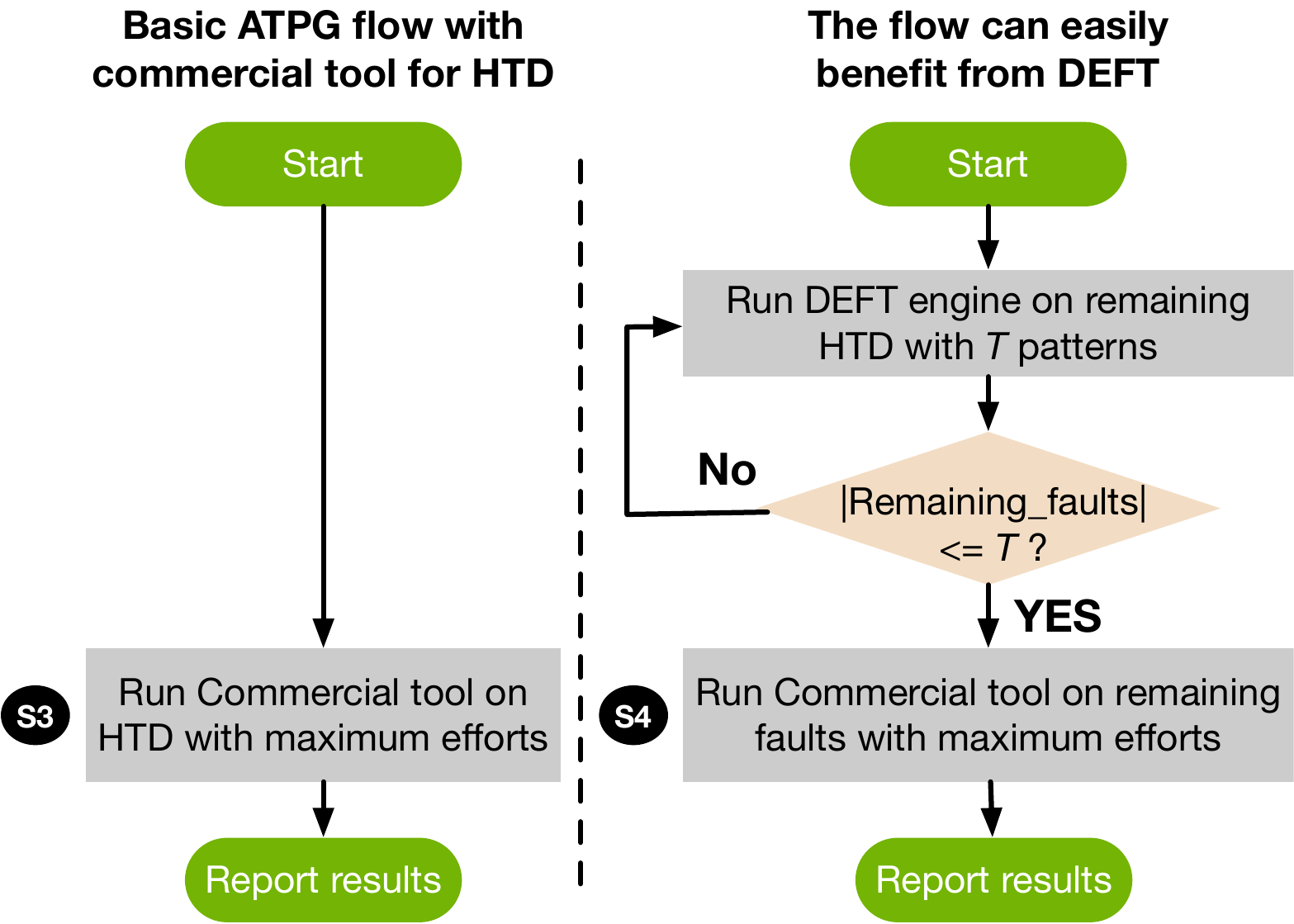}
    \caption{Comparison between the commercial ATPG baseline and a DEFT-assisted ATPG flow on HTD faults.
    % Left: the commercial tool directly targets the HTD set with maximum coverage and compaction efforts.
    % Right: DEFT generates $T$ patterns on the HTD set iteratively, and the commercial tool is then applied to the
    % remaining corner faults with the same maximum-effort setting.
    }
    \label{fig:flow}
\end{figure}

\begin{figure*}[t]
    \centering
    % \captionsetup{skip=3pt,captionskip=0pt}    
    \subfloat[Results on NCU. The commercial tool detects all 348 HTD faults with 116 patterns, \\while DEFT detects all with 96 patterns.]{\includegraphics[height=.285\linewidth]{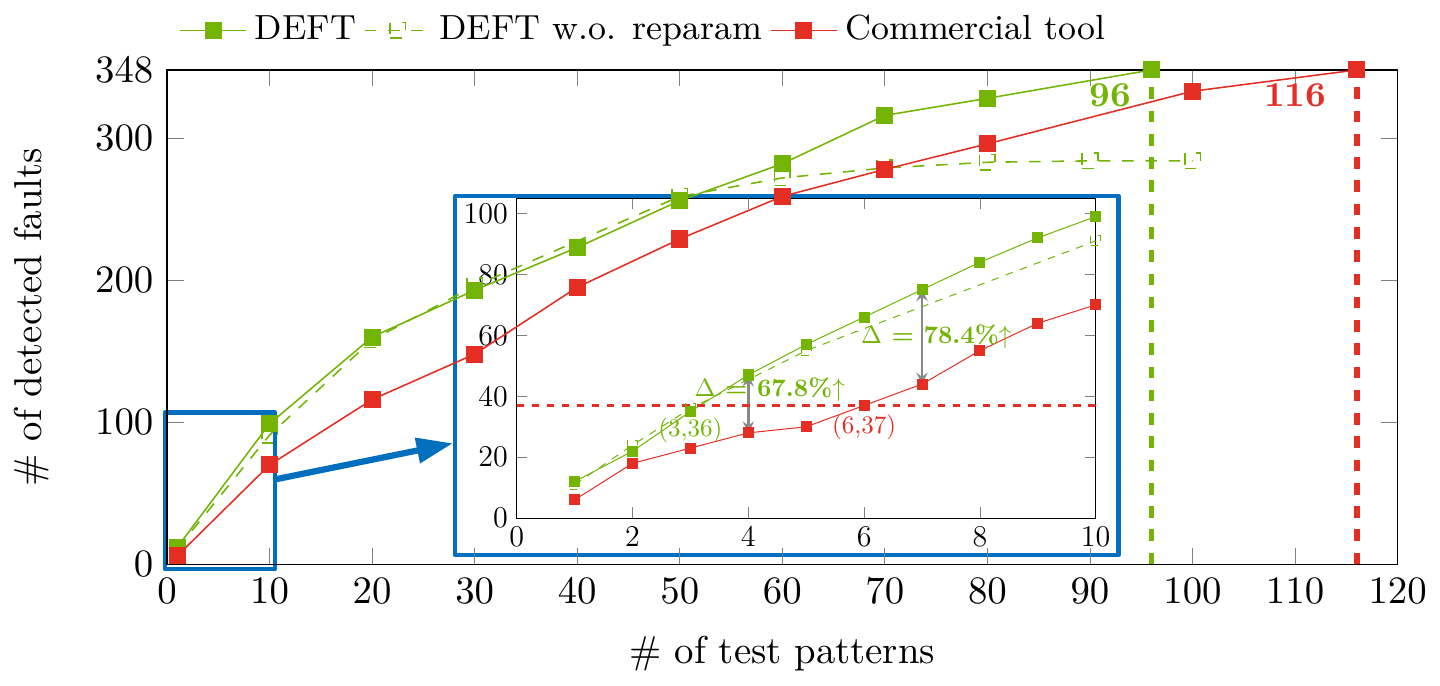}\label{ig:deft:ncu_result}}
    \subfloat[Results on MAC. The commercial tool plateaus after 10 patterns, while DEFT continues to detect additional faults.]{\includegraphics[height=0.27\linewidth]{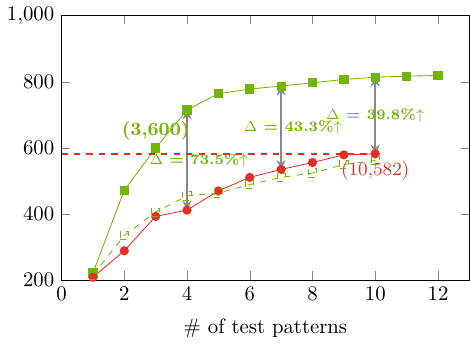}\label{fig:deft:nvdla_result}}
    \caption{ATPG results of DEFT and commercial tool. y-axis is the number of detected faults. }
    \label{fig:result}
    \vspace{-.2cm}
\end{figure*}

DEFT is implemented in Python, using PyTorch for the optimization API and a custom CUDA kernel for efficient forward and backward propagation through the computation graph.
All experiments are conducted on a server equipped with four Intel E5-4620 @ 2.20 GHz CPU cores and a single NVIDIA H200 GPU. 

\new{We evaluate DEFT on the ISCAS'85 combinational suite \cite{iscas85_benchmarks}, the ITC'99 benchmark suite
\cite{itc99_benchmarks}, and two industrial designs: the Non-Cacheable Unit (NCU) from OpenSPARC T2
\cite{opensparc_t2} and a MAC cell from NVDLA \cite{nvdla_open_source}. To focus the comparison on nontrivial
HTD ATPG, we exclude benchmarks for which the commercial tool requires fewer than 10 patterns on the extracted
HTD set. 
% for example, b19 is omitted because the baseline finishes in only three patterns.
 All flip-flops are
treated as scannable PIs/POs, and ATPG is performed on the combinational logic cone. Although DEFT is compatible
with technology-mapped standard-cell netlists, we synthesize all designs into a technology-independent primitive-gate
representation to ensure a uniform evaluation flow. \Cref{tab:benchmark_all} summarizes the benchmark statistics and
HTD ATPG results. 
Overall, the final 0.4\% of faults accounts for about 31\% of the
generated patterns, highlighting why late-stage ATPG dominates pattern growth in practice.
}

% \usepackage{multirow}
\begin{table*}[t]
    \centering
    \setlength{\tabcolsep}{4pt}
    % \hspace*{-0.8cm}
\begin{tabular}{clllllllll|lll|ll} \toprule
\multicolumn{2}{c|}{\multirow{2}{*}{benchmarks}}                & \multicolumn{8}{c|}{benchmark information}                  & \multicolumn{3}{c|}{\# of test patterns} & \multicolumn{2}{c}{runtime (s)} \\
\multicolumn{2}{c|}{}                                           & \# gates & \# PIs & \# POs & \# levels &{\texttt{tp\_limit}}& \# all faults & \# HTD & HTD ratio & Comm.      & DEFT      &  imp.    & Comm. & DEFT \\ \hline
\multirow{7}{*}{ISCAS}             & \multicolumn{1}{c|}{c1355} & 546      & 41     & 32     & 40        & 37             & 3366          & 333    & 9.9\%  & 58                   & 14         & \textcolor{nvidia}{($\downarrow 75.9\%$)} & 0.37            & 3.72    \\
                                   & \multicolumn{1}{c|}{c1908} & 880      & 33     & 25     & 61        & 22             & 4872          & 461    & 9.5\%  & 32                   & 20         & \textcolor{nvidia}{($\downarrow 37.5\%$)} & 0.43            & 12.26   \\
                                   & \multicolumn{1}{c|}{c2670} & 1193     & 157    & 64     & 53        & 33             & 6980          & 680    & 9.7\%  & 47                   & 13         & \textcolor{nvidia}{($\downarrow 72.3\%$)} & 1.27            & 5.96    \\
                                   & \multicolumn{1}{c|}{c3540} & 1669     & 50     & 22     & 71        & 68             & 9360          & 876    & 9.4\%  & 74                   & 18         & \textcolor{nvidia}{($\downarrow 75.7\%$)} & 0.45            & 36.87    \\
                                   & \multicolumn{1}{c|}{c5315} & 2307     & 178    & 123    & 68        & 42             & 13988         & 989    & 7.1\%  & 71                   & 63         & \textcolor{nvidia}{($\downarrow 11.3\%$)} & 0.42            & 86.39   \\
                                   & \multicolumn{1}{c|}{c6288} & 2416     & 32     & 32     & 218       & 8              & 14560         & 999    & 6.9\%  & 19                   & 20         & \textcolor{red}{($\uparrow 5.3\%$)}       & 28.93           & 261.31 \\
                                   & \multicolumn{1}{c|}{c7552} & 2331     & 207    & 108    & 56        & 49             & 14322         & 890    & 6.2\%  & 102                  & 44         & \textcolor{nvidia}{($\downarrow 56.9\%$)} & 1.61            & 83.73    \\\hline
\multirow{17}{*}{ITC99}            & \multicolumn{1}{c|}{b03}   & 222      & 36     & 34     & 20        & 9              & 1312          & 119    & 9.1\%  & 16                   & 9          & \textcolor{nvidia}{($\downarrow 43.8\%$)} & 0.13            & 0.82    \\
                                   & \multicolumn{1}{c|}{b04}   & 654      & 79     & 74     & 33        & 21             & 3778          & 339    & 9.0\%  & 37                   & 22         & \textcolor{nvidia}{($\downarrow 40.5\%$)} & 0.15            & 4.16   \\
                                   & \multicolumn{1}{c|}{b05}   & 730      & 37     & 70     & 63        & 26             & 4120          & 400    & 9.7\%  & 48                   & 26         & \textcolor{nvidia}{($\downarrow 45.8\%$)} & 0.18            & 13.47 \\
                                   & \multicolumn{1}{c|}{b07}   & 581      & 44     & 49     & 37        & 19             & 3262          & 320    & 9.8\%  & 37                   & 23         & \textcolor{nvidia}{($\downarrow 37.8\%$)} & 0.18            & 4.64    \\
                                   & \multicolumn{1}{c|}{b08}   & 228      & 32     & 25     & 33        & 24             & 1312          & 107    & 8.2\%  & 30                   & 23         & \textcolor{nvidia}{($\downarrow 23.3\%$)} & 0.14            & 2.01    \\
                                   & \multicolumn{1}{c|}{b09}   & 240      & 31     & 29     & 23        & 18             & 1354          & 111    & 8.2\%  & 17                   & 17         & -  & 0.13            & 1.49   \\
                                   & \multicolumn{1}{c|}{b10}   & 230      & 30     & 23     & 27        & 18             & 1330          & 125    & 9.4\%  & 23                   & 20         & \textcolor{nvidia}{($\downarrow 13.0\%$)} & 0.13            & 4.58   \\
                                   & \multicolumn{1}{c|}{b11}   & 714      & 39     & 36     & 39        & 47             & 4014          & 364    & 9.1\%  & 65                   & 53         & \textcolor{nvidia}{($\downarrow 18.5\%$)} & 0.16            & 6.53   \\
                                   & \multicolumn{1}{c|}{b12}   & 1400     & 126    & 125    & 34        & 42             & 8000          & 757    & 9.5\%  & 69                   & 47         & \textcolor{nvidia}{($\downarrow 31.9\%$)} & 0.19            & 7.56   \\
                                   & \multicolumn{1}{c|}{b13}   & 414      & 57     & 55     & 22        & 15             & 2364          & 232    & 9.8\%  & 28                   & 21         & \textcolor{nvidia}{($\downarrow 25.0\%$)} & 0.19            & 1.64   \\
                                   & \multicolumn{1}{c|}{b14}   & 4700      & 249   & 269     & 178     & 127             & 26500         & 995    & 3.8\%  & 154                  & 129        & \textcolor{nvidia}{($\downarrow 16.2\%$)} & 0.19            & 70.34   \\
                                   & \multicolumn{1}{c|}{b15}   & 9319     & 453    & 485    & 117       & 335            & 52576         & 991    & 1.9\%  & 179                  & 170        & \textcolor{nvidia}{($\downarrow 5.0\%$)}  & 0.45            & 75.64    \\
                                   & \multicolumn{1}{c|}{b17}   & 29095    & 1350   & 1408   & 161       & 384            & 163544        & 994    & 0.6\%  & 81                   & 76         & \textcolor{nvidia}{($\downarrow 6.2\%$)}  & 56.85           & 340.31    \\
                                   & \multicolumn{1}{c|}{b18}   & 69587    & 2791   & 2776   & 160       & 471            & 388738        & 1311   & 0.3\%  & 23                   & 24         & \textcolor{red}{($\uparrow 4.3\%$)}       & 451.54          & 356.78    \\
%    & \multicolumn{1}{c|}{b19}   & 129750   & 5532   & 5539   & 159       & 501            & 729120        & 1414   & 0.2\%  & 3                    &           &      & 211.10          &      \\
                                   & \multicolumn{1}{c|}{b20}   & 11583    & 463    & 451    & 230       & 266            & 64990         & 998    & 1.5\%  & 150                  & 122        & \textcolor{nvidia}{($\downarrow 18.7\%$)} & 16.55           & 259.93     \\
                                   & \multicolumn{1}{c|}{b21}   & 11871    & 463    & 451    & 235       & 297            & 66608         & 993    & 1.5\%  & 145                  & 97         & \textcolor{nvidia}{($\downarrow 33.1\%$)} & 1.38            & 134.27    \\
                                   & \multicolumn{1}{c|}{b22}   & 17888    & 645    & 633    & 244       & 313            & 100162        & 988    & 1.0\%  & 155                  & 128        & \textcolor{nvidia}{($\downarrow 17.4\%$)} & 9.13            & 126.4   \\\hline
\multirow{2}{*}{Indus.} & \multicolumn{1}{c|}{NCU}   & 112170    & 16584  & 16020  & 72        &     400           &    520898           &     348   & 0.07\%            & 116                  & 96         & \textcolor{nvidia}{($\downarrow 17.2\%$)} & 0.87            & 203.26   \\
                                   & \multicolumn{1}{c|}{MAC}   & 445757   & 6581   & 3595   & 211       &       885         &      2318590         &    1171    & 0.05\%            & 10                   & 3          & \textcolor{nvidia}{($\downarrow 70.0\%$)} & 713.17          & 1056.32    \\\hline
\multicolumn{2}{l|}{Overall} & 728725 & 30788 & 27014 & 2506 & 3976 & 3800900 & 16891 & 0.4\% & 1786 & 1298 & \textcolor{nvidia}{($\downarrow 27.3\%$)} & 1285.19 & 3160.39 \\ \bottomrule
\end{tabular}
\caption{\new{Benchmark statistics and HTD ATPG results for the commercial tool and DEFT. \texttt{tp\_limit} is the
pattern limit used in \protect\scriptlink{app:script-s2}{S2} to define the HTD set, and \textit{imp.} reports the relative
pattern-count change of DEFT with respect to the commercial baseline.}}
\label{tab:benchmark_all}
\vspace{-0.4em}
\end{table*}

\new{\textbf{HTD Fault Set.}
The flow for identifying hard-to-detect (HTD) faults is illustrated in \Cref{fig:htd_process}. 
We first invoke the commercial ATPG tool with its default setting; the corresponding command sequence is
shown in Appendix~\ref{app:commercial-scripts} as \scriptlink{app:script-s1}{S1}. If any faults remain
undetected after this run, including UC (Un-Controlled) or UO (Un-Observed) faults, we directly treat them
as the target HTD fault set. Otherwise, we introduce a test-pattern limit. Specifically, we set
$
k = \min\left(1000, \frac{N_{\mathrm{fault}}}{10}\right),
$
where \(N_{\mathrm{fault}}\) is the total number of faults, and select
\texttt{tp\_limit} as the pattern-count threshold that leaves at most
\(k\) faults undetected. We then rerun the commercial ATPG tool with a pattern limit of \texttt{tp\_limit},
using the script shown in Appendix~\ref{app:commercial-scripts} as \scriptlink{app:script-s2}{S2}. The
remaining undetected faults after this run are defined as the HTD fault set.
}

% We compare DEFT with a leading commercial ATPG tool on hard-to-detect (HTD) stuck-at-fault sets extracted from each benchmark.
% The HTD sets are identified using the commercial tool under its default effort setting.
% For MAC, we collect 1171 remaining undetected faults after a full ATPG run.
% For NCU, we collect 348 remaining undetected faults for NCU after 400 patterns since there is no remaining undetected fault if we execute a full run.

\new{\textbf{Evaluation Flows for Commercial Tool and DEFT.}
Figure~\ref{fig:flow} summarizes both evaluation flows. For the commercial baseline, we run the tool directly
on the HTD set with maximum coverage and compaction efforts, using the script in
Appendix~\ref{app:commercial-scripts} \scriptlink{app:script-s3}{S3}, and report its final results. For the
DEFT-assisted flow, DEFT first targets the same HTD set and generates $T$ patterns. If more than $T$ faults
remain, DEFT continues on the reduced fault set; otherwise, the commercial tool is invoked on the residual
faults with the same maximum-effort setting, as shown in Appendix~\ref{app:commercial-scripts}
\scriptlink{app:script-s4}{S4}
\footnote{Note that \scriptlink{app:script-s4}{S4} is optional; in our experiments, DEFT could detect the remaining faults without the commercial tool.
We include the commercial tool in \scriptlink{app:script-s4}{S4} only to speed up late-stage ATPG. When ATPG is near completion, i.e., each pattern detects only one fault, DEFT cannot further reduce the pattern count.}.
% The maximum iteration of
% DEFT is 1000, and each single DEFT run can be finished within 10 minutes. 
Additional experimental parameters
and implementation details are provided in Appendix~\ref{app:hyperparameters}.}

\new{This experiment demonstrates that DEFT can be inserted into a standard ATPG flow with minimal disruption. At
the same time, the setup is deliberately designed to benchmark the core ATPG engine on the most challenging
faults, since HTD faults dominate test-pattern count in practice and represent a true computational bottleneck
in ATPG. Our goal is to evaluate a new ATPG \textit{paradigm}, rather than a feature-complete, product-level
ATPG \textit{flow}; a full-flow comparison would be dominated by trivial faults and by highly-engineered
features orthogonal to our contribution. We therefore view this integration as an initial deployment path:
with tighter support from existing ATPG infrastructure, such as in-tool fault grouping, DEFT can potentially
evolve beyond the HTD setting studied here and serve as an alternative ATPG engine. 
}

% \begin{table}[]
%     \begin{tabular}{l|llll|ll}
%         \toprule
%         & \# gates  & \# PIs & \# POs & \# levels & test 1\%$^1$ & \# HTD$^2$ \\ \hline
%     NCU & 112170    & 16584  & 16020  & 72        &  49.2\%  &348       \\
%     MAC & 445757      & 6581   & 3595   & 211       & 71.8\%  &1171     \\ \bottomrule
%     \end{tabular}
%     \caption{Benchmark statistics used in experiments.  $^1$`test 1\%' is the percentage of patterns used to detect the final 1\% of stuck-at faults in a full ATPG run by the commercial tool; $^2$`\# HTD' is the number of HTD faults used in experiments.}
%     \label{tab:benchmark}

% \end{table}
\begin{figure*}[t]
    \centering
    \subfloat[Memory comparison. Bars show the maximum feasible batch-pattern combination ($B \times T$).]{\includegraphics[width=0.36\textwidth]{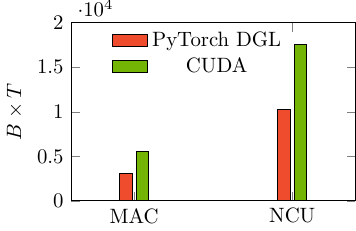}\label{fig:deft:memory}}
    \hspace{0.1\textwidth}
    \subfloat[Runtime comparison on MAC (forward+backward). x-axis is the batch-pattern combination ($B \times T$).]{\includegraphics[width=0.4\textwidth]{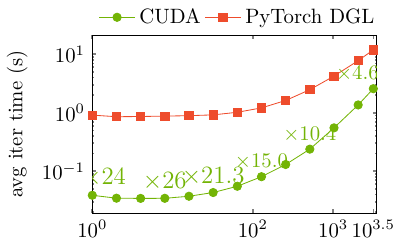}\label{fig:deft:runtime}}
    \caption{Performance comparison of CUDA and PyTorch DGL implementations.}
    \label{fig:cuda_dgl_compare}
\end{figure*}

\begin{figure}[t]
    \centering
    \includegraphics[width=0.4\textwidth]{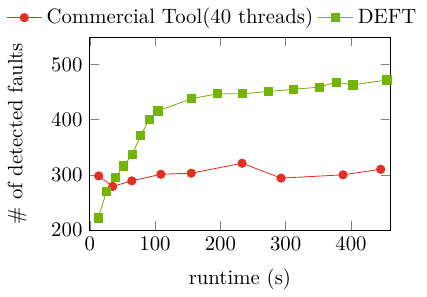}
    \caption{Runtime comparison on MAC with two pattern limits. Commercial tool runtime reflects reported simulation CPU time using 40 threads; its variation comes from different abort-limit settings (200--10,000), while DEFT's variance is from changing the iteration limit (20--700).}
    \label{fig:deft:table5}
\end{figure}

\begin{figure}[t]
    \centering
    \includegraphics[width=0.4\textwidth]{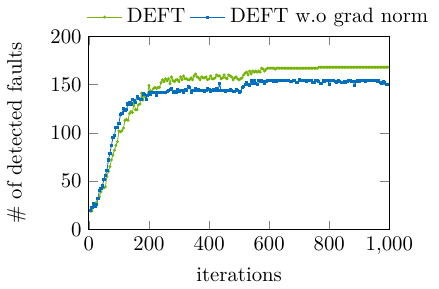}
    \caption{Gradient normalization on NCU with a 20-pattern limit.}
    \label{fig:deft:gradient_norm}
\end{figure}

% \vspace{-0.5em}
\subsection{Benchmark-Wide Results}
\new{\Cref{tab:benchmark_all} shows that DEFT reduces the number of patterns on 23 of the 26 reported benchmarks,
matches the commercial tool on one, and trails slightly on two, yielding an overall 27.3\% reduction. The largest
gains appear on medium-depth designs, where one pattern can still activate and propagate multiple HTD faults, leading
to substantial compaction across both academic and industrial benchmarks. The improvement narrows on deep, structurally
narrow circuits such as c6288, b14, b15, and b17, where each pattern tends to serve fewer HTD faults and the problem
becomes closer to one-fault-at-a-time ATPG. The two losses, c6288 and b18, are both dominated by the final ATPG
endgame; on b18, for example, most HTD faults are redundant, so the task shifts from compacting detections to proving
untestability, which favors the mature commercial engine. Runtime is also not directly comparable, since the commercial
tool is a highly optimized multi-threaded product whereas DEFT optimizes batches of patterns on a GPU. Even so, DEFT
completes every benchmark in under 18 minutes, which is practical for late-stage HTD ATPG.}

\subsection{ATPG Curves on NCU and MAC}
\Cref{fig:result} reports the ATPG results of DEFT and the commercial tool for the NCU and MAC benchmarks.
Across both benchmarks, DEFT consistently achieves higher test effectiveness.
On NCU, it delivers \textbf{67.8-78.4\%} more detected faults in the low-pattern regime and maintains an average 21.1\% advantage.
The margin plateaus because those final faults are exceptionally difficult and cannot be simultaneously excited and propagated by a single pattern. 
Finally, DEFT reaches full coverage with 96 patterns, 17.2\% less than the commercial tool (116 patterns).
In contrast, DEFT without reparameterization (naive relaxation) initially performs comparably but fails on the final HTD faults.

On MAC, DEFT surpasses the commercial tool by detecting 600 faults in just 3 patterns, 
compared to 582 faults in 10. DEFT detects \textbf{48.9\%} more faults on average and ultimately finds over 800 faults, 
40\% of which are missed by the commercial tool. 
This performance contrasts sharply with naive relaxation, underscoring the efficacy of our reparameterization.

\subsection{CUDA Kernel Efficiency}
As shown in \Cref{fig:cuda_dgl_compare}, our CUDA kernel supports about $2\times$ larger batch-pattern capacity ($B \times T$) 
on both benchmarks and delivers \textbf{4.6x-26x} 
lower per-iteration runtime than the PyTorch+DGL baseline on MAC.
The kernel is highly efficient, enabling over 500 patterns to be generated in parallel on MAC, 
with each iteration completing in less than one second.

\subsection{Runtime vs. Quality}
\setlength{\abovecaptionskip}{4pt}

\Cref{fig:deft:table5} compares the runtime-quality tradeoff between DEFT and the commercial tool.
With about 30 seconds, DEFT achieves similar fault coverage as the commercial tool,
and its fault coverage increases rapidly as runtime grows, eventually plateauing above 450 detected faults.
The commercial tool exhibits limited improvement with increasing abort limits, detecting roughly 300 faults at best.
\subsection{X-ratio and Gradient Normalization}
We evaluate DEFT's capability to produce X-bits, as shown in \Cref{tab:xbits},
where $\lambda$ controls the X-bit regularization strength (\Cref{eq:deft:x_penalty}).
Compared with the commercial tool, DEFT achieves higher fault detection across all $\lambda$ settings, improving fault detection by 23.9\%-37.6\%.
Without regularization ($\lambda$=0), DEFT attains a 0/1-bit ratio comparable to the commercial tool, 
indicating that optimization in the relaxed continuous space naturally sharpens only the critical PIs.
As $\lambda$ increases, the 0/1-bit ratio decreases smoothly, confirming that the regularizer effectively promotes X-bit usage.
Even with strong regularization ($\lambda$ = 0.1), \textbf{DEFT retains a 23.9\% advantage in detected faults while reducing the bit-ratio by 30.3\%}.
\Cref{fig:deft:gradient_norm} shows the impact of gradient normalization.
With gradient normalization, the curve converges to a slightly higher fault detection ($>$160), avoiding early saturation at less than 150 faults.

\begin{table}[t]
    \centering
    {    \begin{tabular}{ll|ll} \toprule
        &           & 0/1 bits ratio (\%)& \# detected faults \\ \hline
\multicolumn{2}{l|}{Commercial tool} & 0.393          & 117                \\ \hline
\multirow{4}{*}{DEFT}     & $\lambda = 0$        & 0.413   \textcolor{red}{($\uparrow 5.1\%$)}        & 161  \textbf{\textcolor{nvidia}{($\uparrow$ 37.6\%)}}             \\
        & $\lambda = 0.001$      & 0.338 \textbf{\textcolor{nvidia}{($\downarrow$ 14.0\%)}}           & 160 \textbf{\textcolor{nvidia}{($\uparrow$ 36.7\%)}}               \\
        & $\lambda = 0.01$       & 0.317  \textbf{\textcolor{nvidia}{($\downarrow$ 19.3\%)}}         & 158  \textbf{\textcolor{nvidia}{($\uparrow$ 35.0\%)}}              \\
        & $\lambda = 0.1$         & 0.274  \textbf{\textcolor{nvidia}{($\downarrow$ 30.3\%)}}         & 145  \textbf{\textcolor{nvidia}{($\uparrow$ 23.9\%)}}             \\ \bottomrule
\end{tabular}}
    \caption{X-bit experiment on NCU with a 20-pattern limit. To produce X-bits, DEFT sets $x_i = X$ when $\sigma(\logits_i) \in (0.0001, 0.9999)$. }
    \label{tab:xbits}

\end{table}

\section{Conclusion}
\label{sec:conclu}
This work introduced DEFT, a differentiable ATPG framework that reformulates 
ATPG as a continuous optimization problem and enables scalable gradient-based search on industrial circuits. 
% Through a reparameterized objective and a fused CUDA kernel, 
% DEFT overcomes the semantic mismatch and computational bottlenecks that limit prior relaxation-based approaches. 
Experiments on academic and industrial HTD fault sets demonstrate substantial gains in pattern count and fault detection.
 These results highlight DEFT as a practical and extensible foundation for next-generation ATPG.
%  opening opportunities for integrating differentiable methods into broader design-for-test workflows.
% \section*{Acknowledgments}

\balance
\bibliographystyle{IEEEtran}
\bibliography{ref/floorplan,ref/placement,ref/book,ref/convex,ref/routing,ref/Wei,ref/deft}

\newpage
\appendices
\section{Proofs}
\label{app:proofs}

\subsection{Proof of Theorem \ref{thm:equivalence}}
\label{app:proof-equivalence}

\begin{proof}
Let $\mathcal{X}_D \subseteq \B^n$ be the set of all detecting patterns, i.e.,
$\mathcal{X}_D = \{ \data \in \B^n \mid I_f(\data) = 1 \}$.

\paragraph{P1 solvable $\implies$ P2 optimal value is 1.}
Assume \eqref{eq:p1} is solvable. Then the set of detecting patterns
$\mathcal{X}_D$ is non-empty. Let $\data^* \in \mathcal{X}_D$ be one
detecting pattern. Define $\param^* \in [0,1]^n$ as the deterministic
parameter vector corresponding to $\data^*$:
\begin{equation*}
    \param^*_i =
    \begin{cases}
    1, & \text{if } \data^*_i = 1,\\
    0, & \text{if } \data^*_i = 0.
    \end{cases}
\end{equation*}
For this $\param^*$, the product distribution $p(\data|\param^*)$ is a
Dirac distribution centered at $\data^*$. Therefore,
\begin{align*}
    J(\param^*) &=
    \E_{\data \sim p(\cdot|\param^*)}[I_f(\data)] \\
    &= \sum_{\data \in \B^n} I_f(\data) \cdot p(\data|\param^*) \\
    &= I_f(\data^*) \cdot p(\data^*|\param^*) = 1.
\end{align*}
Since $J_f(\param)$ is the expectation of a binary indicator, it cannot
exceed 1. Hence $\max_{\param \in [0,1]^n} J_f(\param) = 1$.

\paragraph{P2 optimal value is 1 $\implies$ P1 solvable.}
Assume $\max_{\param \in [0,1]^n} J_f(\param) = 1$. Then there exists some
$\param^*$ such that $J(\param^*) = 1$. By definition,
\begin{align*}
    J(\param^*) &=
    \E_{\data \sim p(\cdot|\param^*)}[I_f(\data)] \\
    &= \sum_{\data \in \B^n} I_f(\data) \cdot p(\data|\param^*) \\
    &= \sum_{\data \in \mathcal{X}_D} p(\data|\param^*).
\end{align*}
Thus $J(\param^*)$ is the total probability mass assigned to detecting
patterns. If this mass is 1, then $\mathcal{X}_D$ cannot be empty. Therefore,
at least one detecting pattern exists, and \eqref{eq:p1} is solvable.
\end{proof}

\subsection{Proof of the Logistic Reparameterization}
\label{app:proof-reparam}

\begin{lemma}[Logistic Reparameterization]
\label{lemma:reparam}
Let $\logits \in \R^n$ be a vector of logits. Let $\noise \in \R^n$ be a
noise vector with independent components
$\noise_i \sim \text{Logistic}(0,1)$, drawn by sampling
$u_i \sim \mathcal{U}(0,1)$ and setting
$\noise_i = \log(u_i) - \log(1-u_i)$. Define
\begin{equation}
    \xh(\logits,\noise)_i = \I(\logits_i + \noise_i > 0).
\end{equation}
Then $P(\xh(\logits,\noise)_i = 1) = \sigma(\logits_i)$.
\end{lemma}

\begin{proof}
By definition,
\begin{align*}
    P((\xh)_i = 1)
    &= P(\logits_i + \noise_i > 0) \\
    &= P(\noise_i > -\logits_i).
\end{align*}
The standard Logistic distribution has cumulative distribution function
$F(y) = P(\noise_i \le y) = \frac{1}{1+e^{-y}} = \sigma(y)$. Therefore,
\begin{align*}
    P(\noise_i > -\logits_i)
    &= 1 - F(-\logits_i) \\
    &= 1 - \sigma(-\logits_i) \\
    &= \frac{e^{\logits_i}}{1+e^{\logits_i}}
     = \frac{1}{1+e^{-\logits_i}}
     = \sigma(\logits_i).
\end{align*}
\end{proof}

\section{DEFT Hyperparameters}
\label{app:hyperparameters}

\begin{table}[h]
    \centering
    \caption{Key hyperparameter settings used in all DEFT experiments.}
    \label{tab:hparams}
    \begin{tabular}{p{0.28\columnwidth} p{0.62\columnwidth}}
    \toprule
    \textbf{Hyperparameter} & \textbf{Setting and Description} \\
    \midrule
    \textbf{Gumbel temperature}
    & $\tau$ is annealed from $\tau_{\text{start}}=3.0$ to
    $\tau_{\text{end}}=0.5$ using a cosine schedule. A high initial
    temperature encourages exploration in the relaxed space, and the low
    final temperature sharpens the distribution for discrete decisions. \\ \hline
    \textbf{Explore weight}
    & $w_{\text{explore}}$ is annealed linearly from $w_{\text{start}}=1.0$
    to $w_{\text{end}}=0$ over the entire run. This provides strong
    early-stage guidance from continuous surrogates while ensuring that the
    final optimization is governed by the exploit objective. \\ \hline
    \textbf{Learning rate and optimizer}
    & A constant learning rate of $0.1$ is used for the AdamW optimizer.
    This value balances training stability and convergence speed across all
    benchmarks. \\ \hline
    \textbf{Monte Carlo samples}
    & Each batch entry draws $K_{\text{samples}}=3$ Gumbel perturbations.
    Final evaluation uses $K_{\text{eval}} = 3 \times \text{batch\_size}$
    samples. In our experiments, $K_{\text{samples}}$ has no significant
    impact on result quality. \\ \hline
    \textbf{Batch size}
    & Let $B$ denote the maximum batch size given the available memory and
    computational resources. We set the batch size as $B/4$. We note that number above $B/4$ will linearly increase the runtime without significant improvement in result quality. \\ \hline
    \textbf{Epochs}
    & Each DEFT run executes for 1000 epochs. This was sufficient for all
    faults to converge under the annealing schedule. \\ \hline
    \textbf{Multi-fault and multi-pattern}
    & DEFT optimizes over the entire HTD fault set in each run. For MAC, $T=1$.
    For others, $T=40$.
    We note that the convergence speed is strongly affected by the solution space, determined by the
    circuit scale, the number of faults, and the pattern budget. 
    We leave the exploration of more advanced multi-fault and multi-pattern optimization strategies to future work.
    \\ \hline
      \\
    \bottomrule
    \end{tabular}
\end{table}

\section{Commercial ATPG Scripts}
\label{app:commercial-scripts}

The command snippets below summarize the key commercial-tool scripts used in the experimental flow.\footnote{For
simplicity, prerequisite setup commands such as \texttt{read\_verilog} are omitted.}

\medskip
\noindent\hypertarget{app:script-s1}{\scriptbadge{S1}}\hspace{0.5em}\textbf{Default run for HTD identification}
\begin{quote}\small\ttfamily
\detokenize{add_faults -all}\par
\detokenize{create_patterns}
\end{quote}

\medskip
\noindent\hypertarget{app:script-s2}{\scriptbadge{S2}}\hspace{0.5em}\textbf{Pattern-limited run for HTD identification}
\begin{quote}\small\ttfamily
\detokenize{add_faults -all}\par
\detokenize{create_patterns -Pattern_count ${tp_limit}}
\end{quote}

\medskip
\noindent\hypertarget{app:script-s3}{\scriptbadge{S3}}\hspace{0.5em}\textbf{Commercial baseline on the HTD set}
\begin{quote}\small\ttfamily
\detokenize{read_faults ${HTD}}\par
\detokenize{set_abort_limit 500}\footnote{We note that increasing abort limit does not show an obvious pattern count decrease. See \Cref{fig:deft:runtime}.}\par
\detokenize{create_patterns -COVerage_effort high -COMpaction_effort Maximum}
\end{quote}

\medskip
\noindent\hypertarget{app:script-s4}{\scriptbadge{S4}}\hspace{0.5em}\textbf{Commercial finishing step after DEFT}
\begin{quote}\small\ttfamily
\detokenize{read_faults ${final_remaining_faults}}\par
\detokenize{set_abort_limit 500}\par
\detokenize{create_patterns -COVerage_effort high -COMpaction_effort Maximum}
\end{quote}

\end{document}